\documentclass[preprint]{aastex}
\usepackage{color, soul}
\usepackage{graphics}
\usepackage{subfigure}
\usepackage{rotating}
\usepackage{lscape}
\usepackage{cases}
\usepackage{gensymb}
\usepackage{changepage}
\begin{document}

\title{A Deep {\it Chandra} View of a Candidate Parsec-Scale Jet from the Galactic Center Super-massive Black Hole}
\author{Zhenlin Zhu\altaffilmark{1,2}, Zhiyuan Li\altaffilmark{1,2}, Mark R. Morris\altaffilmark{3}, Shuo Zhang\altaffilmark{4}, Siming Liu\altaffilmark{5}}
\affil{$^{1}$ School of Astronomy and Space Science, Nanjing University, Nanjing 210023, China}
\affil{$^{2}$ Key Laboratory of Modern Astronomy and Astrophysics (Nanjing University), Ministry of Education, Nanjing 210023, China}
\affil{$^{3}$ Department of Physics and Astronomy, University of California, Los Angeles, CA 90095, USA}
\affil{$^{4}$ MIT Kavli Institute for Astrophysics and Space Research,
Cambridge, MA 02139, USA}
%\affil{$^{5}$ Columbia Astrophysics Laboratory, Columbia University, New York, NY 10027, USA}
\affil{$^{5}$ Purple Mountain Observatory, Chinese Academy of Science, Nanjing 210034, China}
\email{zhuzl@smail.nju.edu.cn; lizy@nju.edu.cn}
\begin{abstract}
We have investigated the linear X-ray filament, G359.944-0.052, previously identified as a likely X-ray counterpart of a parsec-scale jet from the Galactic Center super-massive black hole (SMBH), Sagittarius A* (Sgr A*), using a total of $\sim$5.6 Ms ultra-deep {\it Chandra} observations taken from September 1999 to July 2017. 
This unprecedented dataset enables us to examine flux and spectral variations that might be related to intrinsic properties of the weakly accreting SMBH.  
We find no flux or spectral variation in G359.944-0.052 after the G2 periapsis passage around early 2014, however, {  a moderate flux increase of $\sim$2\,$\sigma$ significance} might be associated with the periapsis passage of G1 in early 2001. 
The filament exhibits an unusually hard spectrum (photon-index $\lesssim$1) in its portion closest to Sgr A* (i.e., near-side) and a significant spectral softening in the more distant portion, which can be interpreted as synchrotron cooling of the relativistic electrons moving along the jet path.
In particular, the hard spectrum of the near-side suggests a piling up of quasi-monoenergetic electrons caused by rapid radiative cooling. 
{  The spectral and temporal properties of G359.944-0.052 strengthen the case of it being the X-ray counterpart of a jet launched by Sgr A*.}
\end{abstract}

\section{Introduction}
\label{sec:intro}
%%% BACKGROUND
Accretion onto a super-massive black hole (SMBH) can produce highly collimated, magnetized outflows of relativistic particles, i.e., jets. 
While the launching mechanism and composition of jets are still not well understood, it is generally believed that jets can mediate the transport of an enormous amount of energy from the {\it central engine} to much greater physical scales, thus playing an important role in regulating the co-evolution of the SMBH and its environment {  (\citealp{Meier2012})}.
As such, jets often manifest themselves as elongated features across a broad range of wavelengths, especially in the radio and X-ray bands where synchrotron and/or inverse Compton radiation from relativistic electrons are predominant.
Spatially-resolved studies of such features have greatly facilitated our general understanding of jet energetics and kinematics.

\par

%%% LMB13 Intro
Rather ironically, our knowledge about jets emanating from the closest SMBH, the one located in the Galactic center (GC) and best known as Sgr A* \citep{Melia2001}, remains elusive. 
On the one hand, numerous theoretical studies have demonstrated that the centimeter-to-millimeter emission from Sgr A* can well be interpreted as synchrotron radiation from a relativistic jet (e.g., \citealp{Falcke1993}; \citealp{Falcke2000}; \citealp{Markoff2001}; \citealp{Yuan2002}), which is likely symbiotic with a radiatively inefficient, advection-dominated accretion flow (\citealp{Yuan2014}).
On the other hand, despite the continuing effort, observational searches for the putative jet, on sub-pc to kpc scales and over radio to $\gamma$-ray wavelengths, remain inconclusive {  (for an overview of the proposed jet candidates, see Li, Morris \& Baganoff 2013; hereafter LMB13; see also \citealp{Sha2015}; \citealp{Yusef2016}). }

Among the proposed jet manifestations, a narrow linear X-ray feature named G359.944-0.052 (hereafter G359.944 for brevity), originally identified by \citet{Muno2008} with {\it Chandra} observations and further studied by LMB13, is of special interest for several reasons.
First, this feature traces the putative jet on a sub-pc scale, 
%which is well within the gravitational influence radius of Sgr A*, 
{  thus it can potentially help constrain any jet-driven feedback in the close vicinity of Sgr A*}, as well as reveal short-term variations in the accretion process. 
Second, this feature is detected in X-rays and its power-law-shaped spectrum was found to be consistent with synchrotron emission. Due to the strongly magnetized environment of the GC, the synchrotron cooling timescale, on the order of $\sim$1 yr (see discussion in {  Section 6}), requires continuous energy input, which was suggested to originate from the putative jet dissipating its internal energy upon collision with one of the gas streamers constituting {  the so-called {  mini-spiral} (Figure 1), first discovered by \citet{Lo1983} and \citet{Ekers1983} with the Very Large Array (VLA). }
Third, and perhaps physically most attractive, the inferred spatial orientation of G359.944 (hence that of the underlying jet) {  is aligned }with the angular momentum of the Galactic disk.  This alignment is what one might expect if: 1) the jet orientation is dictated by the SMBH's spin axis, and 2) if the SMBH's angular momentum has resulted from the accretion of stars and gas that had an average angular momentum reflecting that of the Galaxy.
This special geometry, if true, is highly instructive for modeling the process of accretion onto Sgr A*.  
Indeed, recent studies combining numerical models and millimeter data seem to agree on a high inclination angle (with respect to our line-of-sight) of the spin axis (e.g., \citealp{Vincent2015, Broderick2016}), lending support to the inferred jet path. 

 \par

%%%Motivation for revisit
{  The case for a jet can be reinforced} if a relation between the variability of G359.944 and intrinsic variability of Sgr A* could be established. 
It has long been known that Sgr A* exhibits strong variability in its broadband radiation, a phenomenon commonly dubbed {\it flares} (\citealp{Baganoff2001}; \citealp{Genzel2003, Ghez2004}).
%(i.e., strong and rapid flux variations) in the X-ray, infrared and radio bands. 
In particular, the X-ray flares of Sgr A* can reach peak fluxes $\sim$10--100 times the quiescent level on a timescale of minutes (e.g., \citealp{Baganoff2001,Porquet2003,Nowak2012,Zhang2017}), strongly suggesting that they arise from merely a few gravitational radii from Sgr A*. %the inner region of the accretion flow.
At present, there is no consensus on the physical origin of the flares, but it is reasonable to associate flares with fluctuations or instabilities in the magnetized accretion flow (e.g., \citealp{Chan2015, Ball2016, Li2017, Yuan2018}). 
%In addition, as the Galactic Center gas cloud/star G2 reaches its closest approach with the central SMBH SgrA* (Burkert et al. 2012; Gillessen et al. 2013; Witzel et al. 2014; Pfuhl et al. 2015), in spite of the non-detection of significant luminosity variation, the possible variation of accretion rate might be reflected in the jet luminosity.
For this reason, when the G2 object, a hypothetical dusty cloud (with or without a central star; \citealp{Gillessen2012,Witzel2014}), was discovered to be moving on a highly eccentric orbit ($e \approx$ 0.98) and approaching periapsis around early 2014,
there was heated anticipation that it might boost accretion onto Sgr A* and trigger strong radiation. %is on a highly eccentric orbit ($e \approx$ 0.98) with a pericenter radius of roughly 20 light hours from Sgr A* (\citealp{Gillessen2012} {  [quote one more reference, say,from the UCLA group, which holds the central-star view]}).
%(Burkert et al. 2012; Gillessen et al. 2013; Witzel et al. 2014; Pfuhl et al. 2015; Ponti et al. 2015),
Indeed, an increase in the frequency of bright X-ray flares a few months after the periapsis passage of G2 was suggested by \citet{Ponti2015} (see also \citealp{Mossoux2017}). 
On the other hand, no significant excess was seen at other wavelengths after the periapsis passage (For radio, see \citealp{Park2015, Tsuboi2015}; For infrared, see \citealp{Witzel2018}; For $\gamma$-rays, see \citealp{Ahnen2017}). 
%On the other hand, no evidence was found in the radio band for asymmetric outflows/jets after the periapsis passage (\citealp{Park2015}).
\par
{  The jet power is generally thought to be correlated with the accretion rate (\citealp{Yuan2014}).}
In this regard, a parsec-scale jet may provide additional insight to the accretion process that is possibly affected by tidal interaction between Sgr A* and G2 or similar objects (e.g., G1; \citealp{Witzel2017}).
In this work, we aim to revisit the case of G359.944 being the X-ray counterpart of the Sgr A* jet, taking advantage of the extensive {\it Chandra} observations of the GC in the past two decades.
Our focus is devoted to probing flux and spectral variations in G359.944, and to further constraining the X-ray radiation mechanism and underlying jet properties.   

We describe the {\it Chandra} data in Section \ref{sec:data}. 
The spatial properties of G359.944 are reviewed in Section \ref{sec:xray}.
Analysis of the flux variability is presented in Section \ref{sec:timing}.
In Section \ref{sec:spec}, we examine the spatially-resolved and time-dependent spectra of G359.944. 
Implications for the radiation mechanism and jet properties are discussed in Section \ref{sec:dis}, followed by a summary in Section \ref{sec:sum}.
We adopt a distance of 8 kpc for the GC (\citealp{Ghez2008, Gillessen2009}).

\section{Data Preparation}
\label{sec:data}
The inner few parsecs centered on Sgr A* have been frequently visited by the {\it Chandra X-ray Observatory} since launch, chiefly with its Advanced CCD Imaging Spectrometer (ACIS).
In this work, we utilize 47 ACIS-I observations taken between September 1999 and March 2011,
38 ACIS-S observations with the High Energy Transmission Grating (HETG) taken between February and October 2012, and 39 ACIS-S non-grating observations taken between May 2013 and July 2017.
All these observations had their aimpoint placed within $1'$ from Sgr A*\footnote{Two ACIS-I observations (ObsID 14941 and 14942) taken in 2013 also satisfy this criterion. These two ObsIDs are not included here, since we wish to have a clear temporal division between the three datasets.}, thus ensuring an optimal point-spread function for narrow features like G359.944.
%chiefly to resolve the accretion flow onto Sgr A* (\citealp{Nowak2012}; \citealp{Wang2013}).
The ACIS-I data are essentially the same as used in LMB13.   
The ACIS-S non-grating (hereafter referred to as ACIS-S) observations, while taken in a 1/8 subarray mode, have a sufficiently large field-of-view to cover G359.944 and serve to substantially extend the temporal baseline, especially around the periapsis passage of G2. 

\par
Our data reduction procedure is detailed in Zhu, Li \& Morris (2018), in which we have combined the ACIS-I and ACIS-S/HETG (hereafter simply referred to as HETG) observations to obtain the deepest ever X-ray source catalog of the GC.
%\citet{Zhu2018}. 
The main steps are briefly described below.
We uniformly reprocessed the level 1 event files with CIAO v4.9 and the corresponding calibration files, following the standard pipeline\footnote{http://cxc.harvard.edu/ciao}. 
%We used the CIAO tool {\it acis\_process\_events} to reprocess the level 1 event files and {\it reproject\_aspect} to calibrate the relative astrometry among the individual observations.
The light curve of each ObsID was examined, and when necessary, was filtered to remove time intervals contaminated by significant particle flares.
This resulted in a total cleaned exposure of 1.42 Ms from ACIS-I, 2.83 Ms from HETG and 1.33 Ms from ACIS-S.
The three datasets have comparable sensitivities\footnote{With the HETG inserted, about half of the incident X-rays are dispersed, while the remaining X-rays continue to the detector directly and form the ``zeroth-order" image.}, and when combined, they roughly double the signal-to-noise ratio (S/N) as achieved in LMB13 for G359.944.
It is noteworthy that two ACIS-I observations, ObsID 5360 and 6639, suffer from relatively high particle background throughout their short exposures, thus they were excluded from the following analysis. 
In total, 122 observations spanning 5.58 Ms are included in this work. More specific information on the datasets are listed in Table \ref{tab:log}.
After the cleaned level 2 event file was created for each ObsID, we produced a merged event file by reprojecting all events to a common tangential point, i.e., the position of Sgr A* ([RA, DEC]=[17:45:40.038, -29:00:28.07] at epoch J2000).
We also generated exposure maps in the 2--8 keV band for each observation, assuming an incident spectrum of an absorbed bremsstrahlung. The lower energy cutoff is justified by the large foreground absorption column density $N_{\rm H}\sim10^{23}{\rm~cm^{-2}}$ (e.g., \citealp{Zhu2018}).   
The individual exposure maps of the same instrument (i.e., ACIS-I, HETG or ACIS-S) were then reprojected to form a combined exposure map. 

%The 2-8 keV counts image created from the merged event file is shown in Figure \ref{fig:image}, marked with the position of Sgr A* and the apex defined in LMB13.
%and found that only mild particle flares were present in $\lesssim$1\% of the total time intervals, which have a negligible contribution ($\lesssim 5\times10^{-4}$) to the total background. Hence we decided to preserve all the science exposures for source detection and characterization, maximizing the useful signals.
{ 
In addition, we have acquired ancillary VLA and {\it NuSTAR} images, chiefly to obtain constraints on the flux of G359.944 at radio frequencies and in hard X-rays. 
Details of the VLA images can be found in Zhao et al.~(2009; see also LMB13).
%To provide a constraint on the high-energy X-ray ($>$10 keV) emission from the jet candidate, 
We analyzed the {\it NuSTAR} data obtained in 2012 (obsID: 30002001001, 30002001003 and 30002001004), an epoch with no X-ray transient contaminating the inner few parsecs around Sgr A* (S. Zhang et al. in preparation). 
%The jet candidate was not detectable in either 3--10 keV or 10--20 keV by NuSTAR as expected, since its X-ray flux as measured by Chandra is below the NuSTAR detection threshold in the crowded Galactic Center region. 
Using the selected dataset, we extracted the source spectrum and the background spectrum from an annular region that can properly represent the PSF contamination from the bright pulsar wind nebula (PWN) candidate G359.95-0.04 \citep{Wang2006} at the location of G359.944. 
We found that G359.944 is detected at only 1\,$\sigma$ confidence level in 3--79 keV, and lower than 1\,$\sigma$ in the 3--10, 10--20 and 20--40 keV energy bands. 
%In 3-10 keV, the corresponding jet flux is F_(3-10 keV)=(1.5+/-1.6)e-14 erg/cm^2/s (error bar is 1-sigma here). While in 10-20 keV, the 1-sigma flux upper limit is 1.9e-14 erg/cm^2/s. 
In 10--40 keV, the 3\,$\sigma$ flux upper limit is 1.5$\times$10$^{-13}$ erg~cm$^{-2}$~s$^{-1}$, assuming a representative photon-index of 1.0 (see Section~\ref{sec:spec}).}
%and \citet{Zhang2017}, respectively. 

\section{Spatial Properties of G359.944-0.052}
\label{sec:xray}
A 2--8 keV counts image combining the 122 {\it Chandra} observations is shown in Figure \ref{fig:image}, highlighting the X-ray filament G359.944 and the X-ray-bright Sgr A complex. Also shown are VLA 8.4 GHz intensity contours tracing the ionized gas streamers of the mini-spiral (\citealp{Zhao2009}).  
The highly linear morphology of G359.944 and its almost perfect alignment with Galactic latitude can be clearly seen in the figure. 
The fact that the extrapolation of the straight line defined by G359.944 points back to Sgr A*, had led \citet{Muno2008} and LMB13 to associate G359.944 with the putative jet. 
LMB13 further identified a shock front on the Eastern Arm of the mini-spiral (at an offset of $[{\Delta}{\alpha}, {\Delta}{\delta}] = [12\farcs8, −7\farcs7]$ from Sgr A*, as marked by the white `X' in Figure~\ref{fig:image}), which they interpreted as the site where the jet penetrates through and dissipates energy to the ionized gas. 
Consequently, G359.944 can be understood as the synchrotron radiation from shock-induced relativistic electrons cooling in a finite post-shock region downstream from the shock along the jet path.
%We can visually claim that the linear filament G359.944 is located along the projection from Sgr A* to the apex (the white `${\rm X}$') identified in the radio image, where the particle acceleration is expected to happen (LMB13).
%It's noteworthy that the orientation of this linear filament is along the Galactic Latitude. 

We revisit the above qualitative picture with the updated X-ray data.
For a quantitative analysis, we follow LMB13 to adopt the hypothetical jet path as pointing away from Sgr A* at a position angle of $124\fdg5$.   
%discernible width of $1\farcs5$. 
%$7.5\arcsec \times1.5\arcsec$ rectangular region (position angle of the long-axis $124\fdg5$) following LMB13 to define G359.944.
Visual examination indicates that this remains an optimal definition for the linear filament in the current X-ray image of significantly enhanced counting statistics. 
%To explore the surface brightness of G359.944, we divide the box region into several smaller ones along the projection and count the 2-8 keV photons in each small box. 
LMB13 suggested that the filament is unresolved along its short-axis.
We update this view by comparing the intensity distribution of G359.944 along its short-axis with that of nearby point sources. 
The derived FWHM is $\sim$$1\farcs1$ for the former and $\sim$$0\farcs8$ for the latter, suggesting that the filament is marginally resolved along its short-axis, having a linear width of $\sim$0.04 pc at the assumed distance of 8 kpc.
{  We emphasize that the average PSF should have a negligible variation along G359.944.}
 
The exposure map corrected 2--8 keV intensity profile along the long-axis of G359.944 is shown in the upper row of Figure \ref{fig:surb_hr}, with the three panels displaying the ACIS-I, HETG and ACIS-S data in order. 
To account for the local background, we have adopted an adjacent box running in parallel with the filament. 
It is clear that the intensity profile looks similar among the three datasets, indicating no drastic changes in the morphology of G359.944 over the past two decades.
According to the presumed physical picture, X-ray emission can be induced immediately downstream the shock front (taken in Figure \ref{fig:surb_hr} as the zero point of the intensity profile). However, as already noted in LMB13 and illustrated in Figure~\ref{fig:image}, the presence of relatively strong and non-uniform diffuse emission around the shock front hampers an accurate determination of the local background thereof. 
To test this possibility, we subtract no background for the first bin ($\lesssim3\arcsec$) of the intensity profile, which thus represents a firm upper limit at the position.
% and is fully consistent with, if not higher than, the mean intensity of the filament starting at $\sim$$3\arcsec$ from the shock front. 
It turns out that this upper limit is compatible with the X-ray filament starting from the shock front. 
In practice, however, we define the filament as starting at $3\arcsec$ and ending at $10\farcs5$ {  (i.e., an apparent length of 7$\farcs$5, corresponding to a physical length of $\sim$ 0.3 pc)}, beyond which point the filament again loses its trace into the diffuse background.
The otherwise smooth intensity profile exhibits a ``knot" at $\sim5\arcsec$, the nature of which is further examined in Section \ref{sec:timing}.
{  For the three individual datasets (ACIS-I, HETG, ACIS-S) and the combined dataset, we find 440/382/504/1326 net counts out of a total (source plus background) of 1049/935/1469/3453 counts, giving a 13.6/12.5/13.1/22.6~$\sigma$ significance to G359.944. }
%In Section \ref{sec:short_var}, we detailedly analyze the short-term variability of this brightest knot and discuss its possible constraint on models.

\par
In the lower row of Figure~\ref{fig:surb_hr}, we present the hardness ratio (HR) profile, where HR $\equiv (I_{4-8 {\rm~keV}}-I_{2-4 {\rm~keV}})/(I_{4-8 {\rm~keV}}+I_{2-4 {\rm~keV}})$, {  and $I$ denotes the intensity of a given band.}
The three datasets again agree with each other, showing an overall trend of softening toward the far-side of the filament {  (a flat HR profile is rejected at 78\%/61\%/92\% confidence level for the ACIS-I/HETG/ACIS-S profiles)}.
As suggested by LMB13, this trend can be understood as synchrotron cooling of the relativistic electrons moving along the jet path.  
%With accumulated exposure, as shown in the lower panel of Figure \ref{fig:surb_hr}, the softening effect is rather clearly indicated at the larger distances.
We can estimate a synchrotron cooling timescale $\tau_{\rm syn} \sim 0.3 (\gamma_em_ec^2/50~{\rm TeV})^{-1} (B/1~{\rm mG})^{-2}$ yr (see Section~\ref{sec:dis} for details), which is comparable to the light crossing time of $\sim$1.3 yr. 
In the meantime, the consistency of the HR profile among the three datasets suggests a stable supply of relativistic electrons at least over the past two decades.
This, however, does not preclude flux and spectral variability on smaller timescales, which will be examined in the following sections.    

\par
%The morphology of G359.944-0.052 resembles a PWN candidate to some extent.
%We further compare the surface brightness of this filament along the minor axis with the averaged surface brightness of point sources located around.(........)

\section{Flux Variation of G359.944-0.052} \label{sec:timing}
%\subsection{Long-term Variability} \label{sec:long_var}
We first probe flux variation in G359.944 by examining the mean 2--8 keV photon flux in all 122 observations ({  individual values are listed in Table~\ref{tab:del_src}}).
The source and background regions are outlined by the two rectangles in the insert of Figure~\ref{fig:image}.
We utilize the CIAO tool {\it aprates}, which applies a Bayesian approach for Poisson statistics in the low-count regime, 
%and Gaussian statistics for moderately large counts, 
to compute the photon flux and bounds for each observation, corrected for the local effective exposure.
For those observations with limited net counts, we derive the 3\,$\sigma$ upper limit. 
The resultant long-term light curve, as shown in Figure \ref{fig:lc}, exhibits no significant inter-observation variability. 
%More quantitatively, we compute the reduced chi-square of these data points, $\chi^{2}/d.o.f. = \frac{1}{N}\sum_{N}\frac{(f-\overline{f})^{2}}{\sigma_{f}^{2}} = 0.99$, which suggests a small scattering from the average value.
{  This is supported by} the {\it normalized excess variance} ({\citealp{Nandra1997, Turner1999}), $\sigma_{\rm rms}^{2} \equiv \sum_{i=1}^N[(f_i-\overline{f})^2-\sigma_{f,i}^2]/(N\overline{f}^2) \approx -0.049$, with 68\% error equaling 0.033, which indicates that the apparent deviations from the mean photon flux, $\overline{f} \approx 1.5\times 10^{-6}~{\rm ph~cm^{-2}~s^{-1}}$, arise predominantly from statistical fluctuations.
%Besides, the delay time of a relativistic jet, $t_{d} = d / c$, where d is the distance from Sgr A* to the filament, is estimated to be $\sim$ 2.2 year. 
%Any influence on the X-ray filament from G2 pericenter passage is expected to be observed starting from the beginning of 2015.
%This result claims the non-detection of any significant long-term variability, particularly after 2015. 
%To reduce the measure uncertainties and further test the non-detection of flux variation, we divide data into several time bins and carry out time-dependent analysis in Section \ref{sec:spec}.

%\subsection{Short-term Variability} \label{sec:short_var}
%However, with the following presented evidence and persuasive arguments in LMB13, we strongly disfavor the PWN scenario for G359.944-0.052. 

%Although LMB13 has presented strong evidence and persuasive arguments to disfavor the PWN scenario for G359.944, we would like to examine this scenario by analyzing the short-term variability of the brightest knot.  
\par
Since the filament has a length of more than one light year, we do not expect to detect intra-observation variability across the whole filament. 
On the other hand, short-term variability might be seen in the ``knot", which is essentially unresolved in the X-ray image of Figure~\ref{fig:image}. 
%located $\sim2\arcsec$ in the near-half the filament, is assumed as the putative pulsar if G359.944 is a PWN.
Therefore, 
%we carry out variability analysis on the brightest knot to see whether it is a viable pulsar candidate.
we search for its variability, employing the CIAO tool {\it glvary} on each of the 122 observations.
This tool uses the Gregory-Loredo variability test algorithm (\citealp{Gregory1992}) on the unbinned X-ray data. 
%using multiple different time bins to look for significant deviation.
We have run the {\it dither\_region} tool to produce a normalized effective area file, 
to be fed to {\it glvary},
%which computes fraction of region area that covers chips or falls onto a bad-pixel.
which accounts for the dead time due to bad pixels or chip gaps. 
This step ensures that any measured time variation is intrinsic and not caused by the telescope dithering. 
%motion of the spacecraft or the pixel-to-pixel variation in the detector effective area.
The output of {\it glvary} assigns a variability index that takes values from 0 to 10,
%with the value of 0 indicating a definitely not variable source and values from 6 to 10 indicating a definitely variable source.
with values $\geq$6 indicating a definitely variable source.
Probable intra-observation variability is only found in one out of the 122 observations, as indicated by its variability index of 6. 
This strongly suggests that the knot maintains a rather stable flux over the past two decades, thus it is unlikely a stellar object. 
%In particular, this reinforces the argument of LMB13 against G359.944 being a pulsar wind nebula (PWN).
%In order to demonstrate whether this variability index is due to real flux variation or simply statistical uncertainty, we have repeated the procedures for a rectangular region in the jet feature, where no short-term variation has been expected.
%As the histogram Figure \ref{fig:var} (b) shown, the distribution of the variability indices is quite similar to the knot, especially noting that there's also another observation measured with variability index = 6.
%Besides, if we consider G359.944-0.052 a PWN, its spectral softening on the far side requires the putative pulsar moving towards SgrA*, rather than away from. The direction of such motion is contradictory to the momentum derived from the shock front (LMB13).

\section{Spectral Properties of G359.944-0.052} \label{sec:spec}
%
%fig1: a ds9 figure of the filament in xray, with arrow and box
%fig2: intensity profile (3figs)
%fig3: hardness ratio (3figs)
%fig4: spectra for the whole filament
%fig5: spectra for the two part of the filament

%table1: observation logs 
%table2 : best-fit parameters of the whole filament spectra in different epochs (jet whole in different epochs; jet near ; jet far )
We now turn to examine the spatially-resolved and time-dependent spectral properties of G359.944.
The spectra are extracted from each ObsID using the CIAO tool {\it specextract}.
For the whole filament, the source and background regions are again defined by the two rectangles in Figure \ref{fig:image}.  
The tool also generates the ancillary response files (ARFs) and redistribution matrix files (RMFs), weighted over the source region. 
%The background-extraction region is chosen to be the dashed rectangle in the vicinity of G359.944, denoted with dashed line in insert image of Figure \ref{fig:image}.
We then produce an average spectrum of ACIS-I, HETG or ACIS-S, by combining ObsIDs taken with the same instrument and weighting the ARFs and RMFs by the effective exposure.  
We note that although the instrumental response of {\it Chandra} has degraded over the years of its operation, the effective area of a given instrument above $\sim$2 keV (i.e., the energy range of interest) has undergone no significant change, which justifies the analysis of the combined spectra.
Throughout this section we report errors at 90\% confidence level unless otherwise noted.

%Therefore, we separately merge G359.944 spectra extracted from data of the three different instruments into three combined spectra.
{  All fitted spectra are adaptively binned to achieve S/N better than 3 per bin over the range of 1--9 keV.} 
As shown in Figure~\ref{fig:spec}a, all three spectra appear featureless (i.e., consistent with non-thermal emission) and can be well-fitted with an absorbed power-law model, {\it tbabs*powerlaw} in {  XSPEC v12.9.1} (\citealp{Wilms2000}).
It is reasonable to assume that the absorption column density toward the inner 30$\arcsec$ around Sgr A* does not significantly vary on a timescale of two decades, thus we apply a joint fit to the three spectra, linking the column density but allowing the photon-index to vary. This yields $N_{\rm H} = 17.6^{+5.1}_{-4.2}\times10^{22}~{\rm cm}^{-2}$, consistent with the typical value derived from X-ray point sources in this region (\citealp{Zhu2018}). 
%best value of which we then use for the following spatially or temporally resolved spectral fitting.
The best-fit photon-index ($\Gamma_1$) is largest ($1.40^{+0.50}_{-0.50}$) in the ACIS-I spectrum and smallest ($0.62^{+0.58}_{-0.60}$) in the ACIS-S spectrum, {  suggesting the two values differ at 90\% significance.}
%although the uncertainty is substantial in each case. 
In the meantime, a possible small increase in the unabsorbed 2--10 keV luminosity is found in the ACIS-S spectrum, by $\sim$$(30\pm20)\%$ as compared to the ACIS-I spectrum (Column 6 of Table \ref{tab:spatial}).
The observed spectrum might have been manipulated by foreground dust scattering, which results in spectral hardening due to an $E^{-2}$ dependence in the scattering cross-section \citep{Predehl1995}. 
%To account for this effect, we introduce optical depth $\tau_{sca}$: $\tau_{sca} = 0.087\times A_{v}({\rm  mag})\times E({\rm keV})^{-2}$   (\citealp{Predehl1995}), obtaining a column density, $N_{\rm H} = 13.1^{+6.8}_{-4.8}\times10^{22}~{\rm cm}^{-2}$.
We assess this effect under extreme case that the dust scattering is concentrated in a thin plane, obtaining a column density $N_{\rm H} = 13.1^{+6.8}_{-4.8}\times10^{22}~{\rm cm}^{-2}$.
The corresponding best-fit {  photon-indices} ($\Gamma_2$; Column 8 of Table \ref{tab:spatial}) become slightly larger as expected.
However, even under this extreme case, it holds true that the ACIS-S spectrum is rather flat ($\Gamma_2 = 0.83^{+0.58}_{-0.61}$). Below we shall neglect the effect of dust scattering.
\par

%More detailed analysis is necessary to draw further conclusions about spectra of the jet.

%\subsection{Spatial Variation}
%\label{sec:var_spa}
%Roughly estimating from the length of G359.944-0.052, several years might be taken for the injected particles to travel through the entire filament.
As evident in Figure \ref{fig:surb_hr}, the filament exhibits gradual softening towards the far-side.
%In this way, we are also expected to find spectral differences between the near and far halves.
To quantify this softening, we divide the source rectangle into two equal halves and extract a spectrum for each half (Figure~\ref{fig:spec}b-d).
We tie the absorption column densities of both halves to the best-fit value, $N_{\rm H} = 17.6\times10^{22}~{\rm cm}^{-2}$, found for the whole filament. 
This is justified by the fact that K-band foreground extinction across this region varies by less than 10\% (\citealp{Schodel2010}).
We note that fixing the absorption column density effectively eliminates its degeneracy with the photon-index and thus artificially reduces the estimated uncertainty of the latter.
However, it is the relative change in the photon-index, spatially or temporally, that we are most interested in here.
The best-fit photon-indices, listed in rows 4 to 9 of Table \ref{tab:spatial} for the three instruments, {  are different between the near-half and the far-half at more than 95\% confidence level.}
%The difference of the photon-indices, $\Delta \Gamma_{1}$, equals 1.70$\pm$1.06 for ACIS-I, 2.89$\pm$1.16 for HETG and 2.02$\pm$1.26 for ACIS-S.
In particular, the near-half exhibits a very flat spectrum ($\Gamma_1 < 1$) in both HETG and ACIS-S.
It is noteworthy that the knot does not dominate the flux of the near-half, nor does it show a distinct HR (Figure~\ref{fig:surb_hr}). 
The far-half, in contrast, shows a steep spectrum {  consistent with the softening trend seen in the HR profile.}
The unabsorbed 2--10 keV luminosity of the near-half is about 2 times that of the far-half.

%\subsection{Temporal Variation}
%\label{sec:var_tem}
%Utilizing the continuous {\it Chandra} monitoring observation on Sgr A* from 1999 to 2017, we are able to investigate here the possibility of a temporal variation in the luminosity or spectral shape of the jet candidate.
As discussed in Section~\ref{sec:timing}, the statistical uncertainty and the finite light-crossing time may smear any moderate intrinsic variability in G359.944.
In order to reduce the statistical errors, we divide the 18 years of observations into several epochs, requiring at least 300 ks exposure and a minimum baseline of 2 yrs in each epoch. We also ensure that observations from different instruments are not brought into the same epoch. 
As summarized in Table \ref{tab:temporal}, we eventually choose 6 epochs and group the ObsIDs assigned to each epoch to extract a combined spectrum for the whole filament. 
{  5\% of the total exposure from 23 observations are excluded.}
%We aim to compare relative variation among different epochs, therefore, there is no need to take into account the dust scattering when fitting the spectra. 
The fitted results of the combined spectra are shown in Figure~\ref{fig:relation}. 
%The two black curves in Figure \ref{fig:relation}(a) demonstrate the best-fit spectral photon index and luminosity over the past 18 years.
G359.944 reached the softest state ($\Gamma = 1.94^{+0.64}_{-0.67}$) in epoch 2002.2-2002.6, and became harder ($\Gamma = 0.08^{+1.12}_{-1.23}$) in epoch 2008.5-2010.5, {  at $\sim$95\% significance.}
To see whether this apparent difference can be attributed to pure statistical fluctuations {  about a constant power-law spectrum}, we employ {\it multifake} in XSPEC to simulate 1000 spectra for each epoch, feeding the tool with the corresponding ARFs and RMFs and assuming a fixed photon-index of $\Gamma = 1.15$ and unabsorbed 2--10 keV luminosity $L_{2-10} = 2.0\times10^{32}$ erg s$^{-1}$, which represent the long-term mean of the observed spectra. 
%of $2.78\times 10^{-6}$ photons~keV~cm$^{-2}$~s$^{-1}$. 
The fake spectra are then fitted to derive the 90\% percentile of the best-fit parameters, which are taken as the range of statistical fluctuations (the grey strips in Figure~\ref{fig:relation}a). 
The results suggest that there is intrinsic variation in the photon-index of both epochs 2002.2-2002.6 and 2008.5-2010.5. 
%{  It is noteworthy that all known Galactic PWNe show a power-law X-ray spectrum with photon-indices of 1.1-2.2 (\citealp{Kargaltsev2017}). 
%Therefore, the spectra with $\Gamma<1 $ of G359.944 disfavor it being a PWN.}
The apparent variation in $L_{2-10}$, however, can be accounted for by statistical fluctuations. 
Furthermore, as illustrated in Figure \ref{fig:relation}b, there is no significant correlation between $L_{2-10}$ and $\Gamma$, according to the Spearman's rank correlation coefficient r = 0.31, with the p-value of 0.54 for random correlation.

\par
%We further analyze the temporally-resolved spectrum of another X-ray filament named G359.944, a probable pulsar wind nebular (PWN) nearest to Sgr A* (\citealp{Wang2006}; Figure 1), to make a comparison with our jet candidate.
%Following similar procedures, we have extracted the spectrum of G359.95-0.04 from each of the 122 cleaned events files and grouped them into the identical 6 epochs.
%As shown in the upper panel of Figure \ref{fig:relation}a, the photon-indices of G359.95-0.04 show no significant variation around an average value of $\Gamma \approx 1.7$, which is a typical value for PWN X-ray spectrum (\citealp{Gaensler2006}).
%In contrast, the photon-index of G359.944 is on average lower and has larger fluctuation. 

\section{Discussion} \label{sec:dis}
It was anticipated that the periapsis passage of G2, occurring at $T_{0} \sim$2014.2 (marked by a vertical dashed line in Figure \ref{fig:lc}; \citealp{Witzel2017}), might boost accretion onto Sgr A* due to it partial disruption under the strong tidal force of the SMBH. 
However, there seems to be no observational evidence for a dramatic change in the accretion rate, which is presumably manifested by Sgr A*'s broadband radiation, in particular, its quiescent X-ray flux within the {\it Bondi radius} {  (\citealp{Yuan2016,Ahnen2017,Witzel2018})}. 
On the other hand, \citet{Ponti2015} and \citet{Mossoux2017} claimed evidence for an enhanced incidence rate of bright X-ray flares a few months after G2's periapsis passage.
We note that this time delay is comparable to the free-fall time at the periapsis distance of G2 ($\sim$200 AU; \citealp{Witzel2017}).  
If the passage of G2 also had an effect on the putative jet traced by G359.944, we would expect to probe the associated variations after a time delay of $\gtrsim$ 2 yrs, given the distance of G359.944 from Sgr A*.
Thanks to the high-cadence ACIS-S observations since 2013, such an effect can be readily probed.
However, no significant flux variation can be seen when comparing the epochs 2013.5-2014.10 and 2015.5-2017.4 {  (blue and magenta points} in Figure \ref{fig:lc} and Figure \ref{fig:relation}b), assuming that the former epoch represents the normal state of the jet. 
We also examine the light curve near the periapsis passage of G1 at $T_{0}\sim$2001.3. G1 has similar observational properties with G2 in the near-infrared and shows signs of tidal expansion after periapsis passage (\citealp{Witzel2017}).
{  Interestingly, at epoch 2002.2-2002.6, G359.944 exhibits $\sim$2 times flux increase ($\sim2\,\sigma$ significance) as compared to the mean flux before the G1 passage. We caution that a real physical connection is highly uncertain due to the sparse {\it Chandra} observations before and after this epoch.}

  \par
In the proposed jet scenario, relativistic electrons, accelerated at the shock front and streaming down the jet path, are responsible for the observed non-thermal X-ray emission from G359.944.
LMB13 showed that inverse Compton radiation can be ruled out, due to the lack of sufficient seed photons at lower frequencies (e.g., constrained by the upper limits in the radio band as illustrated in Figure~\ref{fig:sed}), which left synchrotron as the most viable radiation mechanism.  
%We further examine the  X-ray spectrum of the immediate downstream region of the shock front.
%However, a nearby gas cloud region with diffuse X-ray emission contaminate the spectrum of the immediate downstream region, resulting in the limited net photons that could not be used to classify it as thermal or non-thermal emission.
Here, we argue that the observed flat spectra, in particular those from the near-half, place strong constraints even on the synchrotron scenario.

According to the standard diffusive shock acceleration (DSA) theory (e.g., \citealp{Drury1983}), post-shock electrons acquire a power-law energy distribution, which can be expressed as: $N(\gamma_{e}) = K\gamma_{e}^{-p}$, where $\gamma_{e}$ is the Lorentz factor and $K$ the normalization factor. 
The slope $p$ is determined by the shock compression ratio $\chi$ as $p = (\chi+2)/(\chi-1)$, while $\chi$ itself can be written as ${\rm \chi} = (\gamma +1)/(\gamma -1 + M^{-2})$ via the adiabatic index $\gamma$ and Mach number $M$.
%We estimate the hardest spectral photon-index that diffusive shock acceleration can produce (\citealp{Drury1983}).
%Shocks in fluids with a ratio of specific heats $\gamma$ have a compression ratio ${\rm r} = \frac{\gamma +1}{\gamma -1 + M^{-2}}$, where M is the shock Mach number.
Therefore, in the case of a strong shock in a non-relativistic plasma, $\gamma = 5/3$ and $M\rightarrow \infty$, we obtain the canonical value of $p =2$, and the corresponding photon-index $\Gamma = (p+1)/2 = 1.5$.
The broadband spectral energy distribution (SED)\footnote{Model calculations and figure plotting are realized with {\it Naima} \citep{Zabalza2015}, a Python package for computation of non-thermal radiation from relativistic particle populations, publicly available at https://naima.readthedocs.io/en/latest/} of such a case is shown as the solid line in Figure~\ref{fig:sed}.
We have assumed an empirical magnetic field strength $B\approx 1 {\rm~mG}$ at the central parsec (\citealp{Plante1995, Eatough2013}), and the minimum/maximum Lorentz factors $\gamma_{e,\rm min}$ = 1 and $\gamma_{e,\rm max}$ = 10$^{8}$. 
%we find that the DSA theory, although being able to account for the non-detection of the radio counterpart for the jet, 
Evidently, this canonical synchrotron cannot account for the flatness of the near-half spectra obtained with ACIS-I, HETG and ACIS-S, with $\Gamma$=$0.92^{+0.53}_{-0.53}$, $0.20^{+0.66}_{-0.68}$ and $-0.10^{+0.63}_{-0.66}$, respectively (see solid lines in Figure \ref{fig:sed}). 
If we further take into account the fact that the strong shock is propagating in the relativistic plasma of the jet, i.e., $\gamma = 4/3$, we can obtain a harder slope of $p=1.5$ and $\Gamma = 1.25$. This latter value, however, still cannot account for the HETG and ACIS-S spectra. 

\par
The flat spectrum motivates us to consider synchrotron from monoenergetic electrons, i.e., represented by a $\delta$-function distribution: $N(\gamma_{e}) = N_{0}\delta (\gamma_{e} - \gamma_{e,0})$ (\citealp{Pacholczyk1970}).
When $\gamma_{e,0}\rightarrow \infty$, the corresponding SED asymptotically approaches $EdN/dE \propto E^{0.3}$, that is, a photon-index $\Gamma = 0.7$, which in principle is the flattest possible synchrotron spectrum. 
This value barely falls within the uncertainties of the photon-indice of the HETG and ACIS-S near-half spectra (Table \ref{tab:spatial}).
In Figure \ref{fig:sed}, we plot the synchrotron spectrum of monoenergetic electrons with an energy of $\gamma_{e,0}m_ec^2=$50 TeV (dashed curve) and $\gamma_{e,0}m_ec^2=$5 TeV (dotted curve), the former approximately matching the observed near-half spectra and the latter consistent with the far-half spectra.
%We might understand the observed X-ray emission as a composition of synchrotron radiation from monoenergetic electrons (e.g., $\gamma_{0}\sim$50 TeV), as a consequence of rapid energy loss at the immediate downstream of the shock front.
We note that these SEDs are fully compatible with the non-detections in radio as well as the 3-$\sigma$ upper limit on the 10--40 keV flux as constrained by {\it NuSTAR}. 
%Considering the synchrotron cooling, the monoenergetic electrons with lower energy (e.g., $\gamma_{0}\sim$5 TeV) exhibit a steeper X-ray spectrum as the dotted line shown in Figure \ref{fig:sed},  which is consistent with the observed far-half spectra of the jet candidate.
%Moreover, DArk Matter Particle Explorer (DAMPE) report a tentative peak at $\sim1.4$ TeV in the observed spectra,  which suggests there are sources ejecting monoenergetic electrons (\citealp{Yuan2017}).
%In sum, we disfavor the diffusive shock acceleration model and further undermine the PWN scenario. 

In reality, a population of quasi-monoenergetic electrons may be generated from rapid synchrotron cooling of an initial population having a power-law energy distribution (\citealp{Rybicki1986}):
%We further investigate this theory by understanding how to produce electrons following $\delta$-function distribution and examining the consistency between synchrotron cooling timescale and the length of the filament.
%Due to the energy loss of synchrotron emission, electrons with initial energy $\gamma_{i}$ cool down to $\gamma$ (Rybicki et al. 1986):
\begin{equation}
N(\gamma_{e})d\gamma_{e} = N_i(\gamma_{e,i})d\gamma_{e,i} \propto \gamma_{e,i}^{-p}d\gamma_{e,i} ,
\end{equation}
\begin{equation}
\frac{d\gamma_{e}}{dt} = -\frac{1}{m_{e}c^{2}}\frac{4}{3} \sigma_{T} c \beta_e^{2} \frac{B^{2}}{8\pi}\gamma_{e}^{2} \equiv -A \gamma_{e}^{2} ,
\end{equation}
where $p > 1$ is the initial power-law slope, $\sigma_{T}$ is the Thomson scattering cross-section, $\beta_e \approx 1$ is the electron velocity relative to the speed of light ($c$). We have $A \approx 1.3\times 10^{-15}(B/1{\rm~mG})^2~{\rm s}^{-1}$.  
From Equations (1) and (2), we can derive the time-dependent electron energy distribution:
\begin{equation}
N(\gamma_{e}) \propto \gamma_{e}^{-p} (1-At\gamma_{e})^{p-2} ,
\end{equation}
where t is the cooling time.
For $1< p < 2$, the electron energy distribution will evolve into a quasi-$\delta$-function peaking at $\gamma_{e} = 1/(At)$ (\citealp{Schlickeiser1984}).
%Therefore, the production of mono-energetic electrons depend on the properties of injected electrons.
We can estimate the time needed for electrons cooling from the near-half (e.g., $\gamma_{e,n}m_ec^2 \approx 50$ TeV) to the far-half (e.g., $\gamma_{e,f}m_ec^2 \approx $ 5 TeV) along the jet path: $\Delta t = 1/(A\gamma_{e,f})-1/(A\gamma_{e,n}) \approx 2.2(B/1~{\rm mG})^{-2}~{\rm yr}$. 
This cooling time is compatible with the half-length of G359.944, $\sim$0.15 pc, if the bulk motion of the electrons is relativistic as expected for a jet. 
We conclude that the above simple model {  is consistent with} the physical picture proposed by LMB13: the putative jet drives a bow-shock in the Eastern Arm and generates ultra-relativistic electrons, which cool by synchrotron radiation and produce the observed X-ray filament downstream the shock along the jet path.  

%\par
%The bolometric luminosity of G359.944 predicted by the above synchrotron model (i.e., $\gamma_{e,0}m_ec^2 \approx 50$ TeV) is $L_{bol} \approx 1.6\times10^{33}$ erg~s$^{-1}$. 
%On the other hand, this luminosity can be expressed as:
%\begin{equation}
%L_{bol} = \int \int m_{e}c^{2}A\gamma_{e}^{2}~N_{0} \delta(\gamma_{e}-\gamma_{e,0}) d\gamma_{e} = \gamma_{e,0}^{2}m_{e}c^{2}AN_{0}.
%\end{equation}
%Therefore, $N_{0} \approx 1.6\times 10^{38}$. We can then compute the total energy of electrons moving in G359.944:
%\begin{equation}
%E_{tot} = \int \gamma_{e}m_{e}c^{2}N_{0} \delta(\gamma_{e}-\gamma_{e,0}) d\gamma_{e} = \gamma_{e,0}m_{e}c^{2}N_{0} \approx 1.3\times10^{40}~{\rm erg}.
%\end{equation}

\section{Summary} \label{sec:sum}
%%%WRITE A SUMMARY HERE
We have utilized $\sim5.6$ Ms of {\it Chandra} observations spanning 18 years to study the X-ray properties of G359.944, a very promising candidate for a jet from Sgr A*. 
Our main results include:
\begin{itemize}
%\item~The theory suggesting G359.944 a PWN is greatly undermined for the following reasons: 1) The highly-collimated and narrow feature. 2) No short-term variability is detected for the brightest knot, which is assumed as the putative pulsar if the theory is correct. 3) The observed hard spectra.
\item~The periapsis passage of G2 in early 2014 does not seem to have caused significant variation in the X-ray spectrum of G359.944. On the other hand, a flux enhancement of {  $\sim$2$\sigma$ significance}, might be associated with the periapsis passage of G1 in early 2001. 
\item~Unusually hard spectra are found in the near-half of G359.944, showing a photon-index as low as $-0.10^{+0.63}_{-0.66}$. The spectrum becomes softer and less luminous further down the putative jet path. Such properties can be best understood as rapid synchrotron cooling of the ultra-relativistic electrons. 

\end{itemize}
%\end{adjustwidth}
We conclude that G359.944 remains a viable candidate for the long-sought jet from Sgr A*. A decisive test may come from imaging of the Event Horizon Telescope available in the near future (\citealp{Psaltis2015}).
Regardless of the validity of G359.944 being the jet, its unusually flat spectrum is intriguing. Previous work has identified more than a dozen filamentary X-ray features within the inner parsecs of the GC (\citealp{Muno2008}; \citealp{Lu2008}; \citealp{Johnson2009}).
Some of these filaments are identified as magnetic flux tubes (\citealp{Zhang2014}; \citealp{Morris2014}), while others might be PWN driven by a fast-moving pulsar (\citealp{Wang2006}; S.~Zhang, et al. in prep.).
{  It is noteworthy that all known Galactic PWNe show a power-law X-ray spectrum with photon-indices of 1.1-2.2 (\citealp{Kargaltsev2017}).}
Six out of the 17 X-ray filaments studied by \citet{Johnson2009} showed a flat spectrum with best-fit photon-index $\lesssim 1.0$ with large uncertainties due to the limited counting statistics. 
It will be an important step to revisit the spectral and temporal properties of these filaments using the updated {\it Chandra} data, which will facilitate a comparison with the case of G359.944 and help us understand the behavior of high-energy particles in the unique GC environment.

\acknowledgements
This work is supported by National Science Foundation of China under grant 11473010. We acknowledge the PIs of the {\it Chandra} programs, in particular Fred Baganoff, for acquiring the data that made this work possible. Z.Z. thanks Xiao Zhang for helpful discussions on the jet model.

\begin{figure*}
     \centering
     \includegraphics[width=\textwidth]{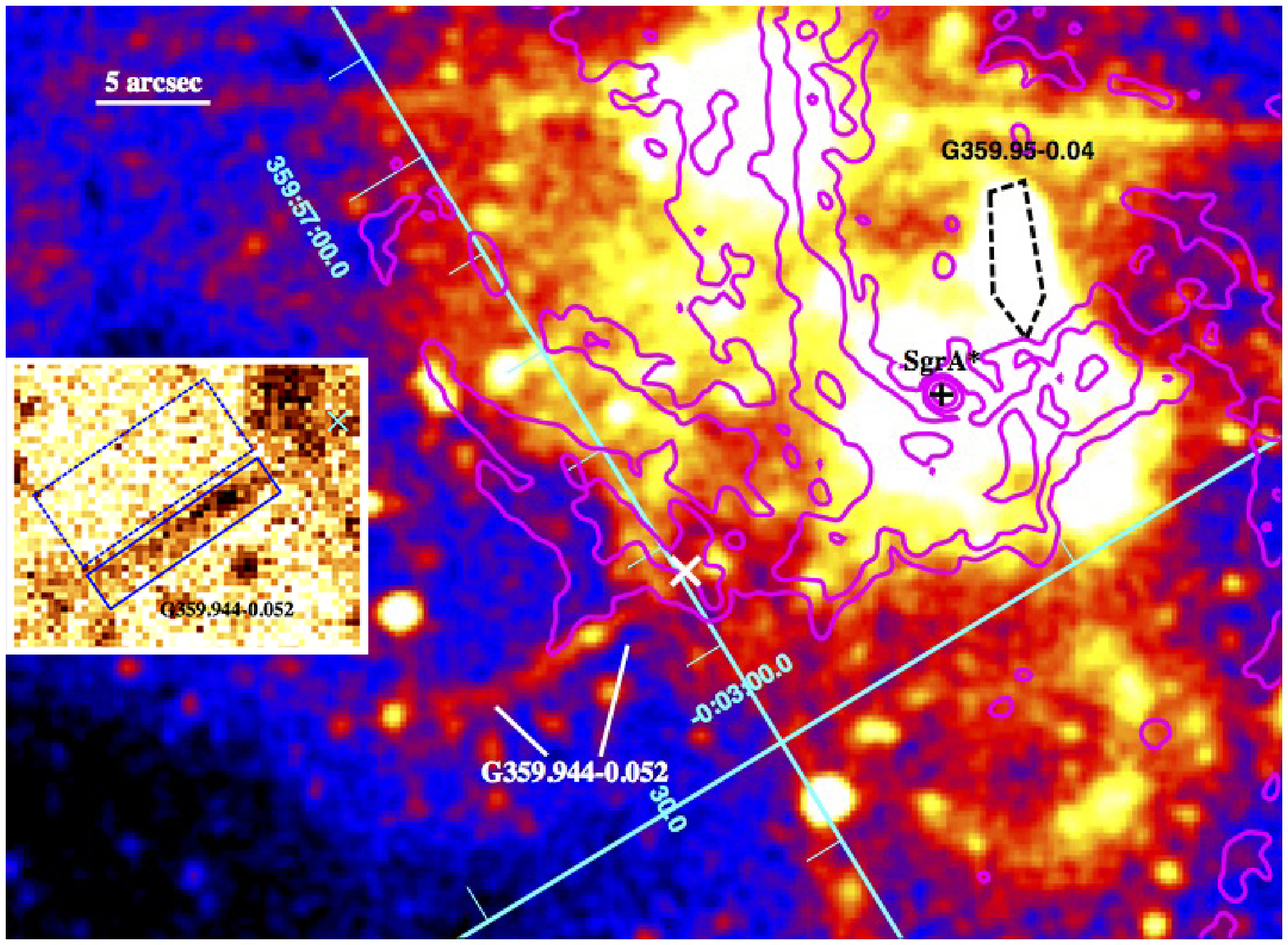}%{fig1_v4.eps}
     \caption{The 2--8 keV {\it Chandra} image of the linear filament G359.944-0.052, {  smoothed with a 2-pixel Gaussian kernel. The cyan grids denote Galactic coordinates.} The `apex' of the shock front (LMB13) is indicated with a white cross. The position of Sgr A* is marked with a `+' sign. The color scale is chosen such that the filament is highlighted while the vicinity of Sgr A* is saturated. {  The magenta contours outline the structure of the mini-spiral as seen in VLA 8.4 GHz intensity map.} We also mark the position of PWN G359.95-0.04 with a dashed polygon. The $7\farcs5\times1\farcs5$ solid box in the insert outlines the source extraction region, and the dashed box ($7\farcs5\times3\farcs0$) denotes the background region. } 
     \label{fig:image}
   \end{figure*}

\begin{figure*}
	\centering	
    \includegraphics[width=\textwidth]{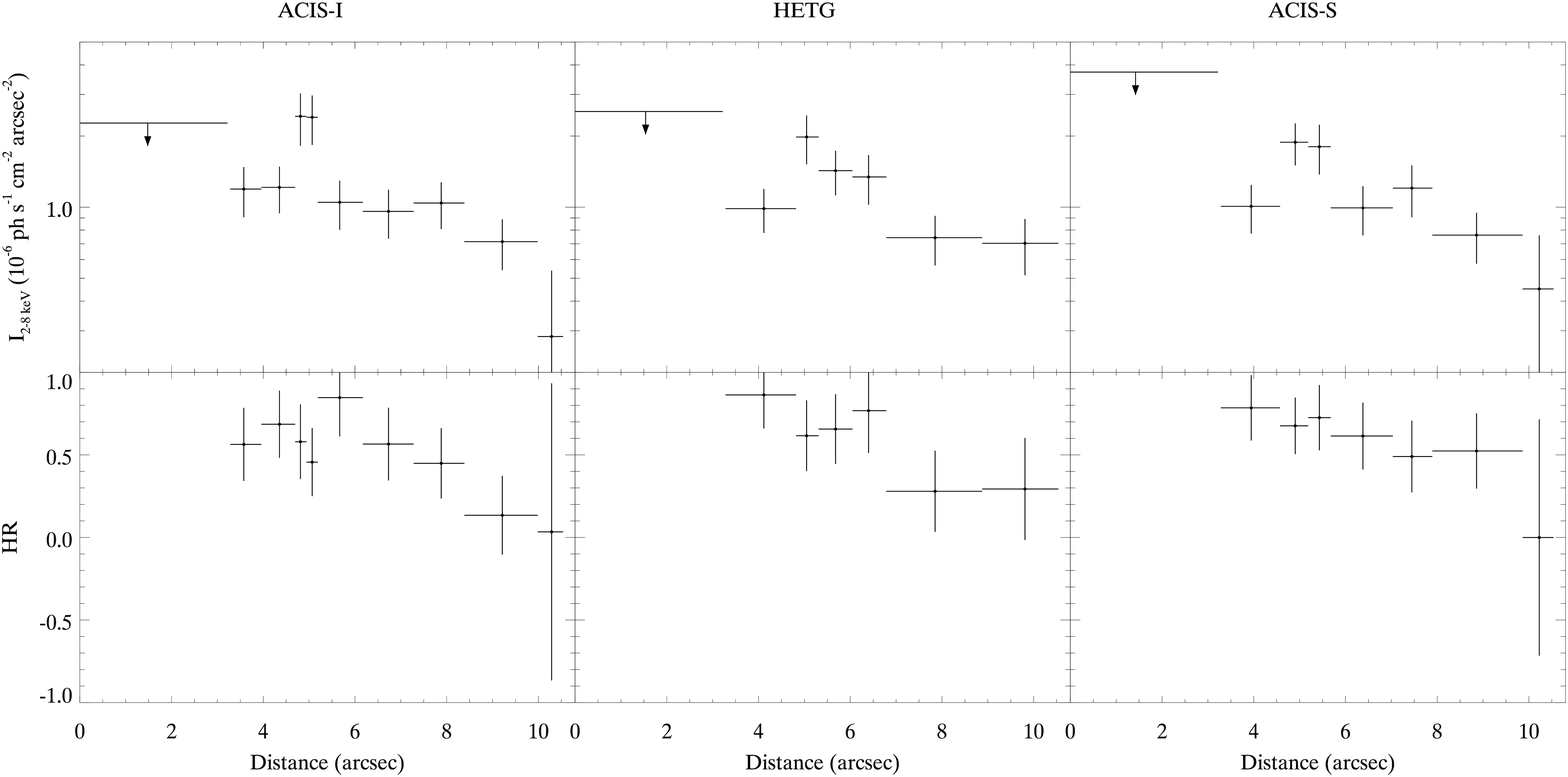}%{fig1_v4.eps}         
      
	\caption{Spatial properties of G359.944-0.052. 
The first row shows the {  2--8 keV intensity ($I_{2-8~{\rm keV}}$)} profiles along the long-axis of the filament, averaged over a width of $1\farcs5$. The three panels display the data of ACIS-I, ACIS-S/HETG and ACIS-S/non-grating, respectively. 
The zero point is at the shock front on the Eastern Arm (marked by the `X' sign in Figure~\ref{fig:image}).  
The first bin is subject to uncertain local background and thus is considered an upper limit.
The rest of the profile is adaptively binned to achieve a minimum of 40 net counts and a S/N great than 4, {  except for the last bin.}
The second row shows the corresponding hardness ratio profiles, defined as HR $= (I_{4-8 {\rm~keV}}-I_{2-4 {\rm~keV}})/(I_{4-8 {\rm~keV}}+I_{2-4 {\rm~keV}})$. }	
         \label{fig:surb_hr}
\end{figure*}

\begin{figure}
     \centering
       \includegraphics[width=\textwidth]{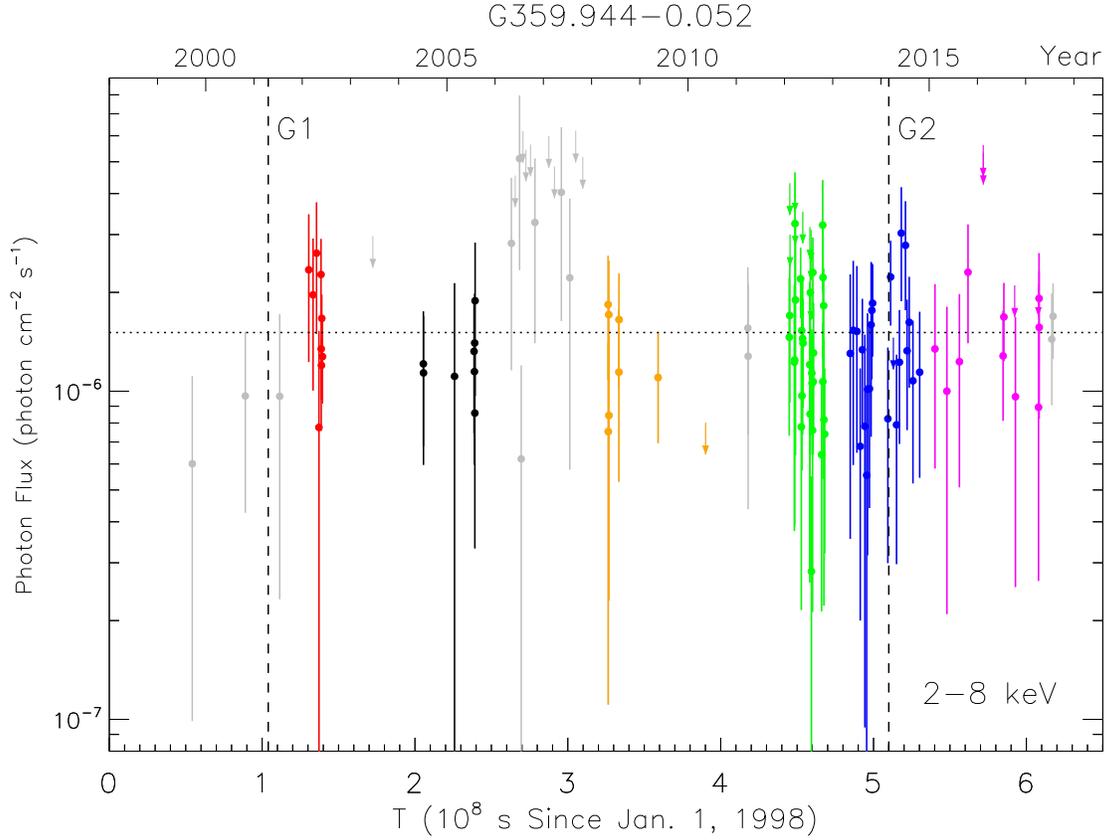}%{jet_lc_2000:8000.eps} 
     \caption{The 2--8 keV light curve of G359.944-0.052, combining the 122 {\it Chandra} observations. Arrows mark 3-$\sigma$ upper limits for observations with limited net counts. Data points of different epochs are color-coded in the same way as in Figure \ref{fig:relation}b. {  Observations not included in the six epochs are marked with grey data points.} The two dashed lines mark the estimated time of closest approach of G1 and G2 (\citealp{Witzel2017}). The mean photon flux is marked as a horizontal dotted line.} 
     \label{fig:lc}
   \end{figure}

\begin{figure}
        \centering
        \includegraphics[width=0.45\textwidth]{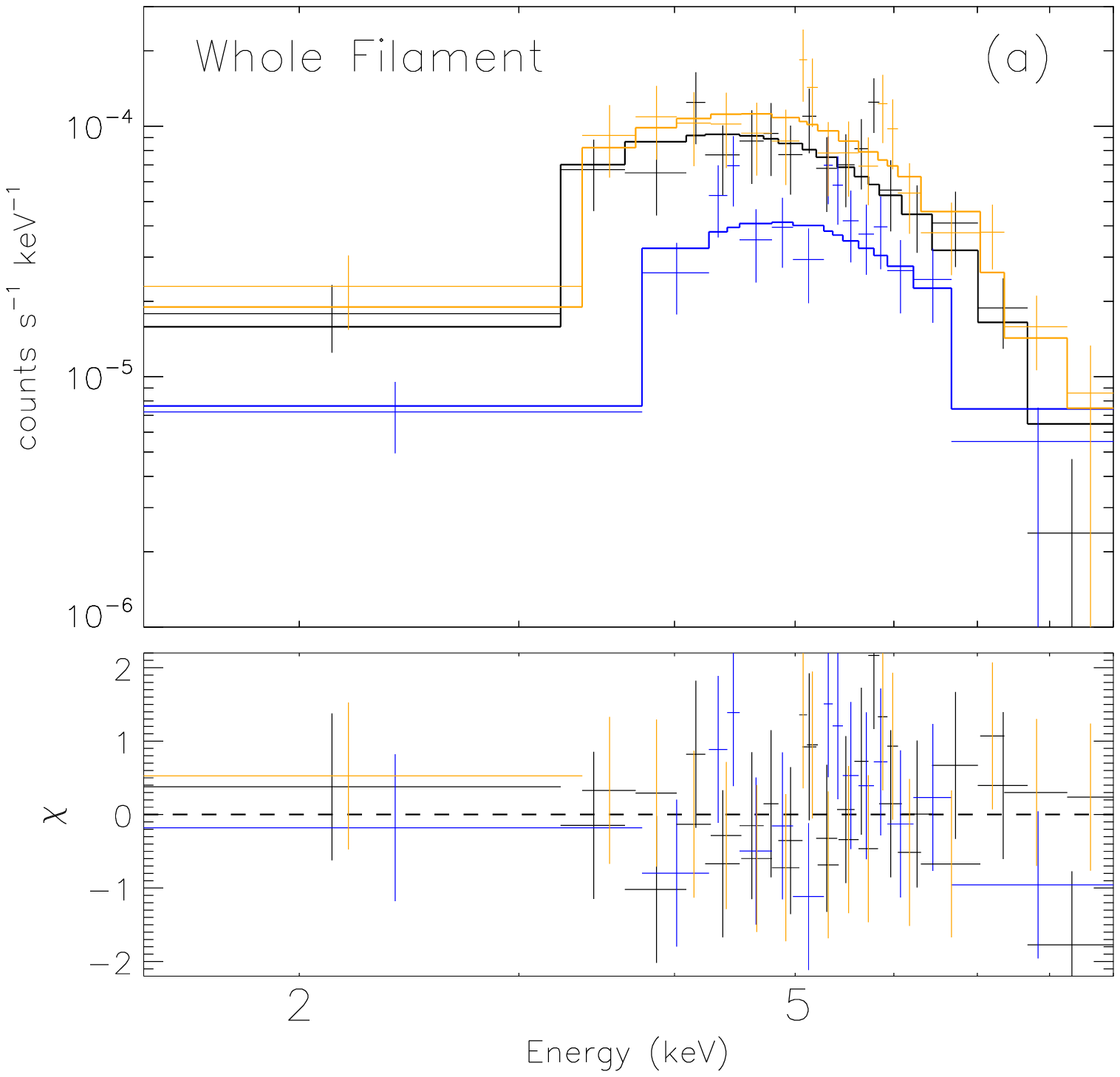}%{spec_jet_all_idl.eps}
	\includegraphics[width=0.45\textwidth]{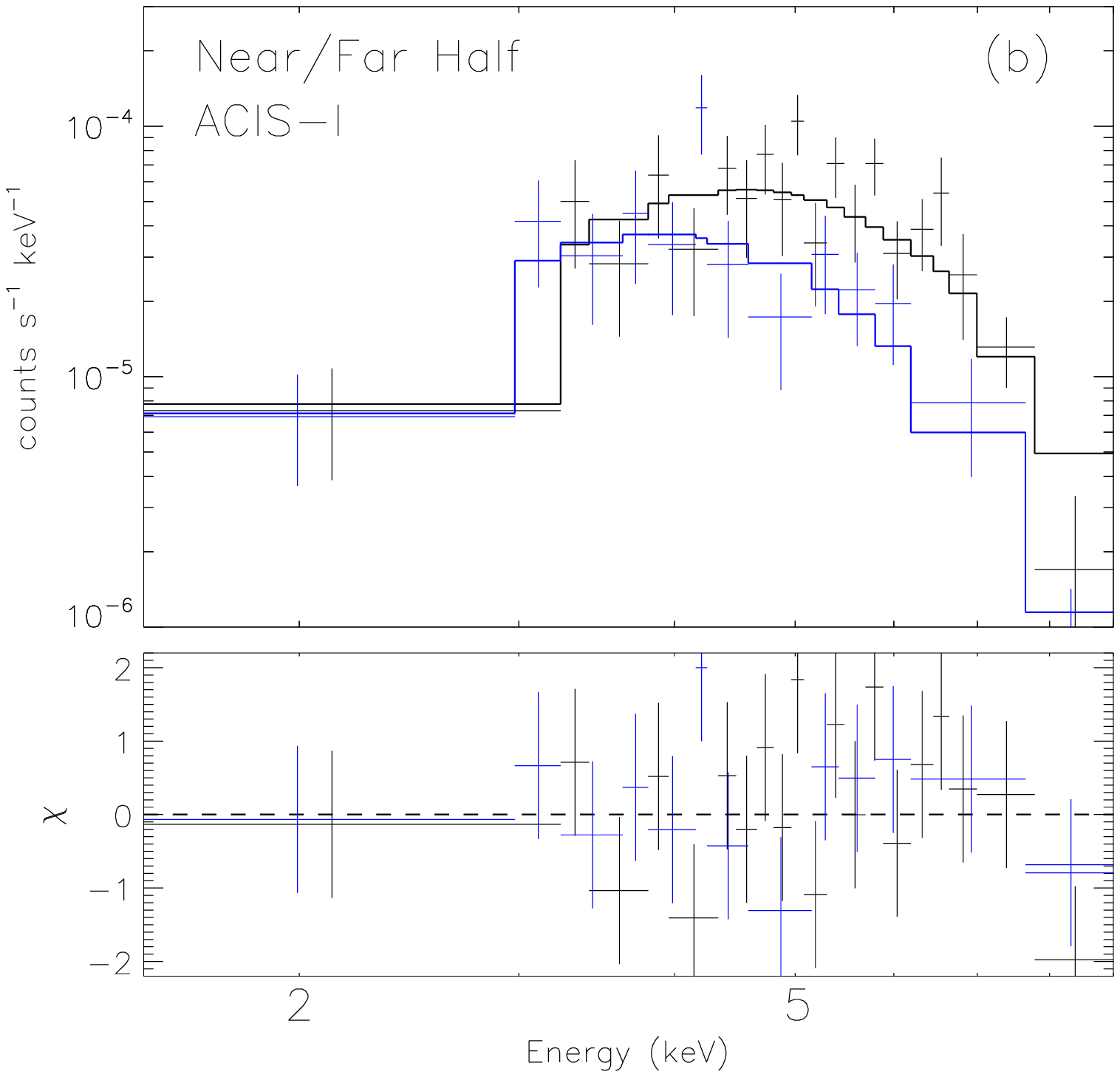}%{jet_near_far_i_idl.eps}
	\\[10pt]
	\includegraphics[width=0.45\textwidth]{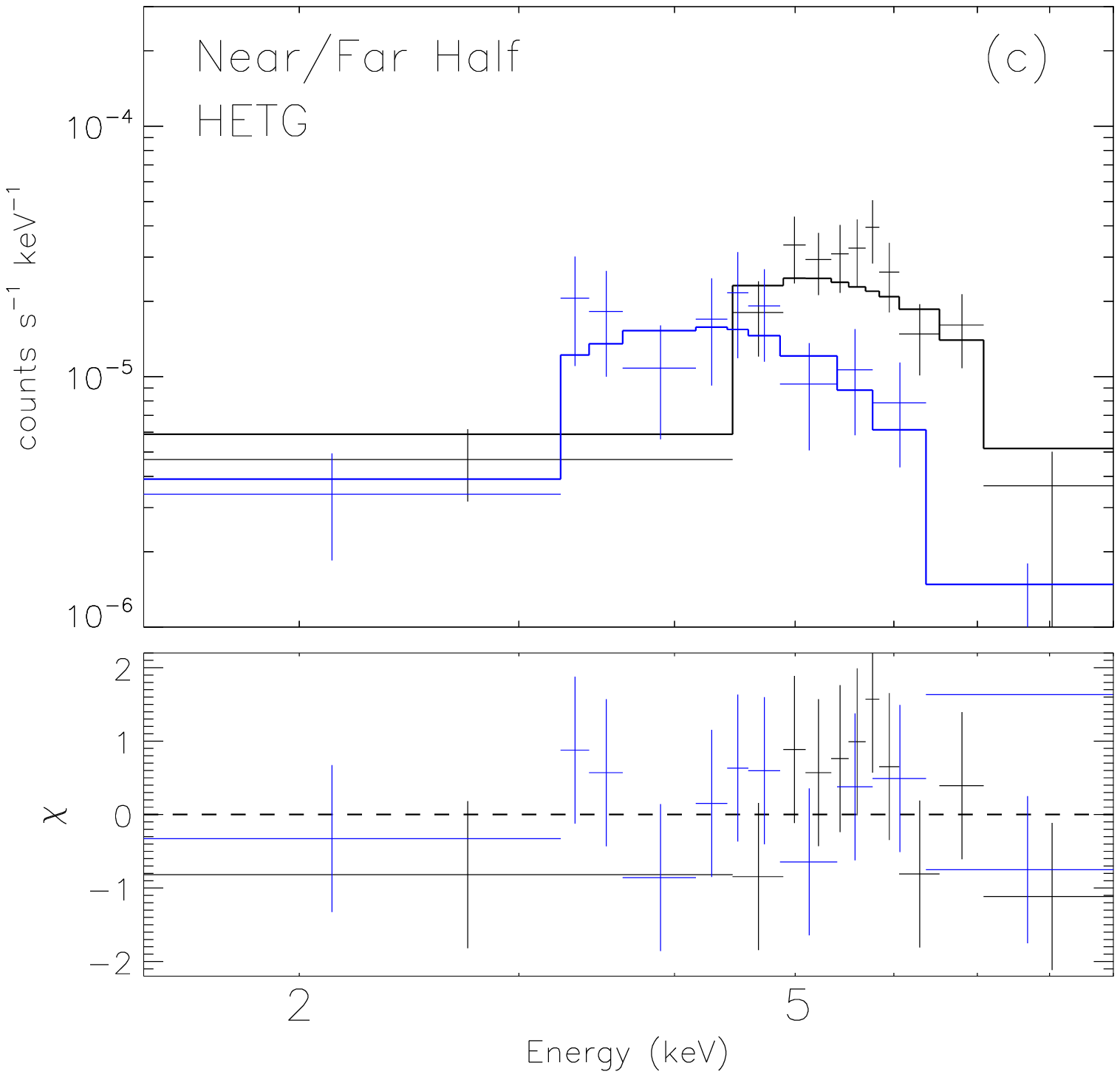}%{jet_near_far_s_idl.eps}
	\includegraphics[width=0.45\textwidth]{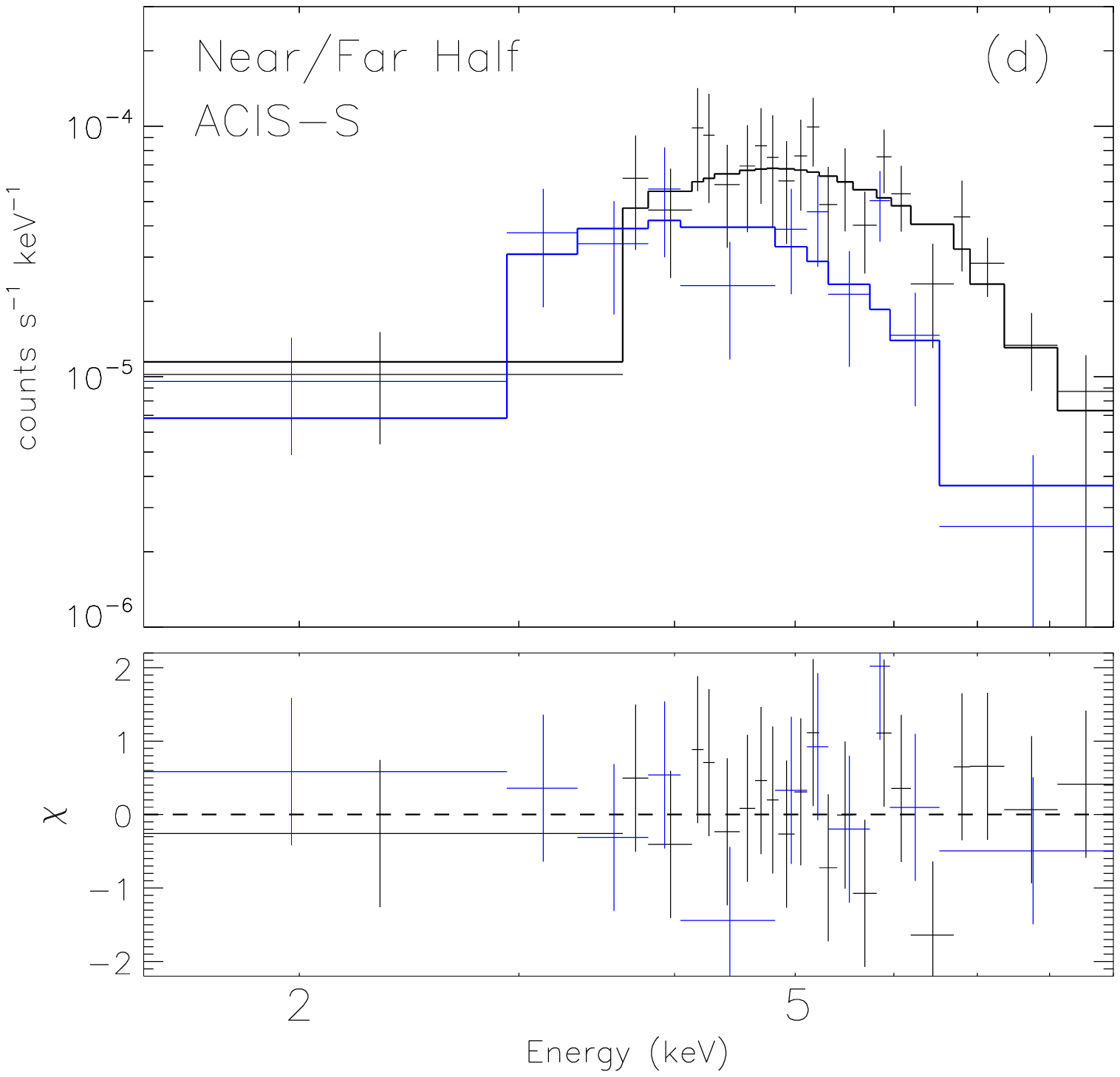}%{jet_near_far_s_none_idl.eps}
        \caption{
        { (a) The combined spectra of G359.944-0.052, adaptively binned to achieve S/N better than 3. The black/blue/orange data points represent the ACIS-I/HETG/ACIS-S spectrum; (b) ACIS-I spectra of the near-half (black) and far-half (blue) of G359.944-0.052. For these two regions, spectra extracted from HETG and ACIS-S are shown in (c) and (d). Also shown in all panels are the best-fit absorbed power-law models. The error bars represent 1-$\sigma$ uncertainty.}
         }	
\label{fig:spec}
\end{figure}

\begin{figure}
     \centering
     \includegraphics[scale=0.57]{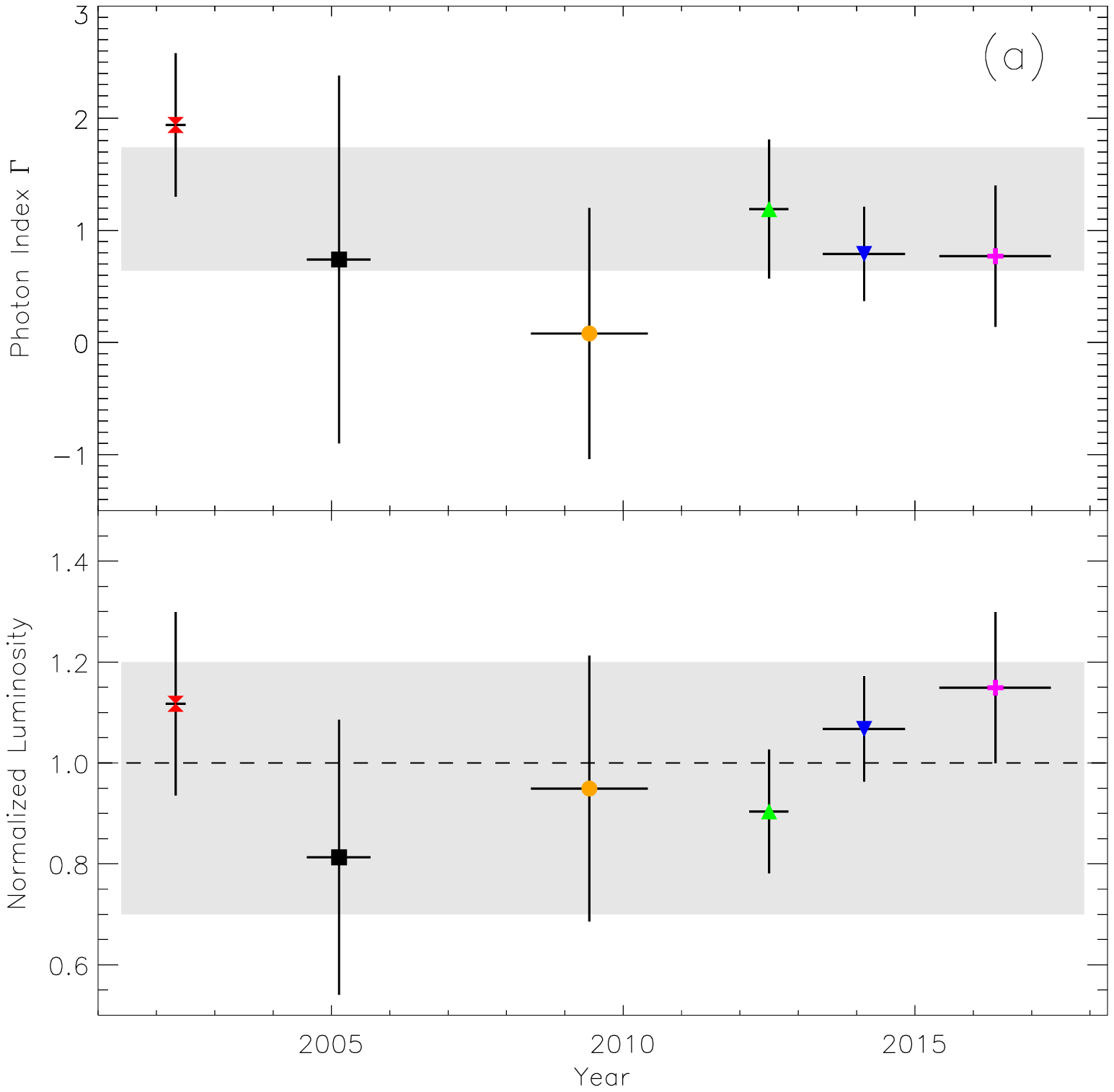}%{epoch_var.eps}
    \\
     \includegraphics[scale=0.57]{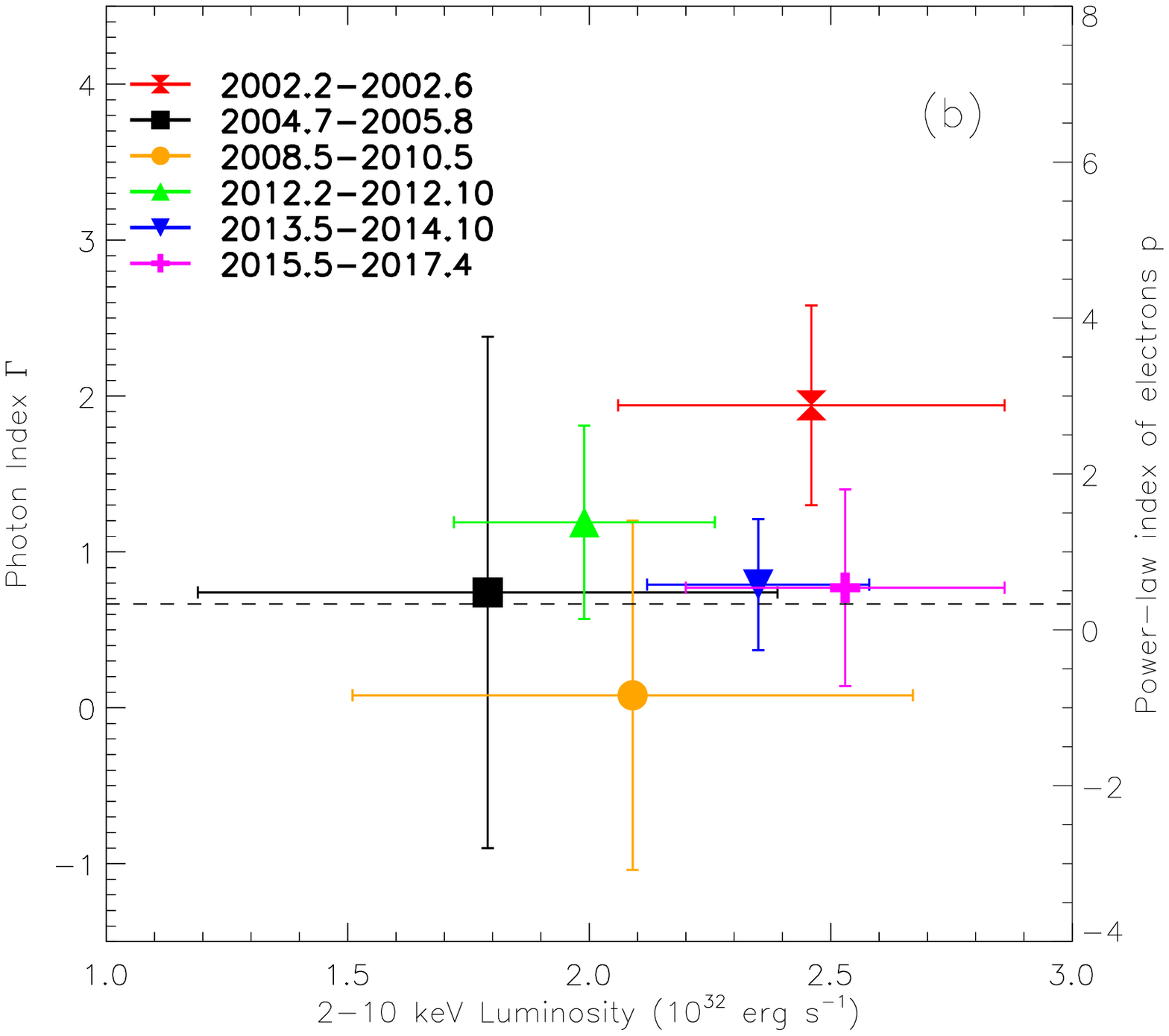}%{epoch_var_par_2axis.eps}
     \caption{(a) Fitted photon-index and 2--10 keV luminosity (relative to the long-term mean) of the whole filament in six epochs. The grey strips indicate the range of pure statistical fluctuations inferred from simulated spectra. 
%The purple curves denote the fitted parameters of the PWN candidate G359.95-0.04.  
(b) The fitted power-law photon index ($\Gamma$; $p = 2\Gamma-1$) versus X-ray luminosity in the six epochs. The dashed line denotes the smallest possible photon-index predicted by the model of monoenergetic electrons (see Section \ref{sec:dis}). The error bars in both panels are of 90\% uncertainty.} 
     \label{fig:relation}
   \end{figure}

% \begin{figure}
%     \centering
%        \includegraphics[scale=0.45]{var_knot.eps}
%           \includegraphics[scale=0.45]{var_near_knot.eps}
%     \caption{(a).Variability indices of the knot feature (putative pulsar in the PWN scenario) measured among 122 observations from 1999 to 2017, the maximum variability index of which is 6. (b). For comparison, the variability indices of a circular region in the jet feature are also derived in the same range of years.} 
%     \label{fig:var}
%   \end{figure}
   
    \begin{figure}
     \centering
        \includegraphics[scale=0.8]{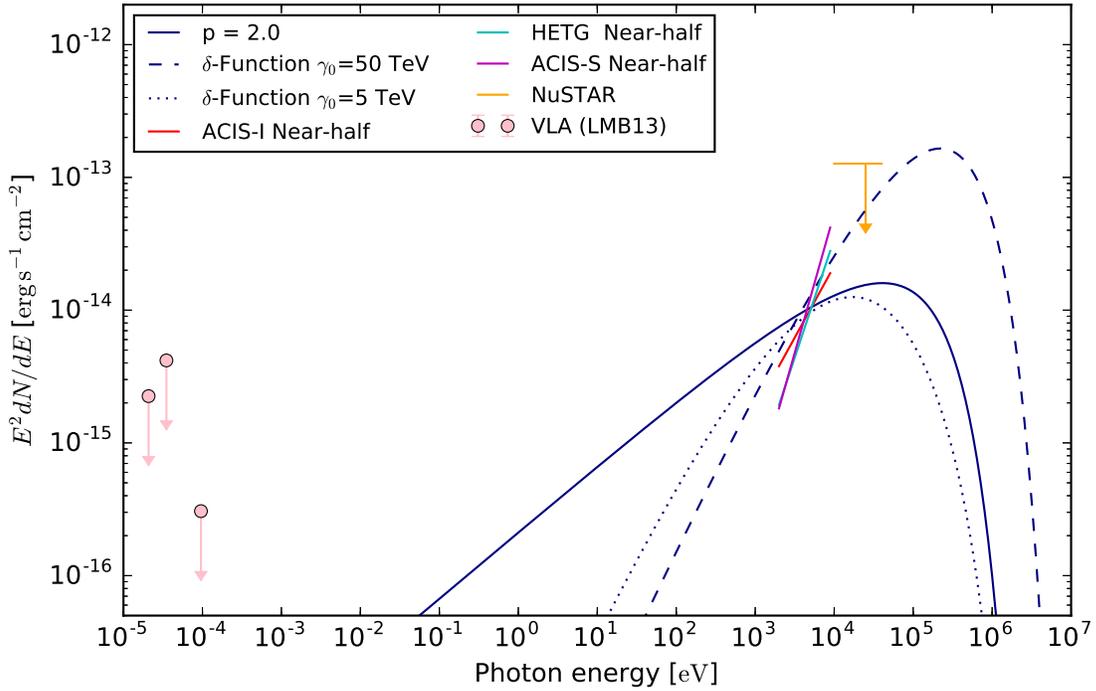}%{jet_sed_all_3sig.eps}
     \caption{Steady-state synchrotron models charactering the X-ray power-law spectrum, which also consistent with current 3-$\sigma$ upper limits of radio data (1.33, 49.8 and 45.0 mJy at 23, 8.4, and 5.0 GHz, respectively; LMB13), denoted with pink arrows. The solid blue curve represents model with p=2.0, where p is the power-law slope of the electron energy distribution. The dashed and dotted curves show synchrotron emission models of monoenergetic electrons, following a $\delta$-function distribution, with $\gamma_{0}$=50 TeV and $\gamma_{0}$=5 TeV respectively. The red, cyan and magenta line segment denotes the averaged near-half spectrum of ACIS-I, HETG and ACIS-S data respectively. The 3-$\sigma$ upper limit given by {\it NuSTAR} in 10--40 keV band is marked with an orange arrow.} 
     \label{fig:sed}
   \end{figure}

%   \begin{figure}
%     \centering
%%     \includegraphics[scale=0.45]{jet_lc_i.eps} 
%%     \includegraphics[scale=0.45]{jet_lc_hetg.eps}
%%     \includegraphics[scale=0.45]{jet_lc_s.eps}
%       \includegraphics[scale=0.4]{jet_lc_2000:4000.eps} 
%       \includegraphics[scale=0.4]{jet_lc_4000:8000.eps} 
%    \caption{2-4 keV vs 4-8 keV lightcurves for jet.}
%   \end{figure}

%%%%%%%%%=========== 	TABLES AND FIGURES    ===============================================
%%%%%%%%%==================================================================================

\clearpage
\begin{deluxetable}{cccccc}
\tabletypesize{\small}
\tablewidth{-50pt}
\tablecaption{{\it Chandra} Observation Log\label{tab:obs}}
\tablehead{
\colhead{ObsID }&
\colhead{Start Time }&
\colhead{Cleaned Exposure}&
\multicolumn{2}{c}{Aim Point}&
\colhead{Roll Angle} \\\cline{4-5}
 & &  & \colhead{R.A. }& \colhead{Dec.}&  \\
  & \colhead{(UT)} &\colhead{(ks)}  & \multicolumn{2}{c}{(J2000)}& \colhead{ (degree) }
}
%Obs.ID &  Start Time & Exposure& Instrument  & \multicolumn{2}{c}{Aim Point}&  Roll Angle \\\cline{5-6}
% &(UT) &  & & R.A. & Dec.&  \\
% & &(ks) & & \multicolumn{2}{c}{(J2000, degrees)}&  (degrees) \\\hline
\startdata
{  ACIS-I}&&&&&\\
242&1999-09-21  02:43:00& 33.3&       266.41399& -29.01271&     268.7\\
15611&2000-10-26  19:08:03& 35.5&       266.41403& -29.01206&     264.7\\
15612&2001-07-14  01:51:10& 13.5&       266.41549& -29.01238&     280.7\\
2951&2002-02-19  14:27:32& 12.4&       266.41862& -29.00345&      91.5\\
2952&2002-03-23  12:25:04& 11.9&       266.41891& -29.00353&      88.2\\
2953&2002-04-19  10:59:43& 11.6&       266.41916& -29.00364&      85.2\\
2954&2002-05-07  09:25:07& 12.4&       266.41938& -29.00374&      82.1\\
2943&2002-05-22  23:19:42& 36.8&       266.41991& -29.00406&      75.5\\
3663&2002-05-24  11:50:13& 37.0&       266.41993& -29.00407&      75.5\\
3392&2002-05-25  15:16:03&164.7&       266.41992& -29.00408&      75.5\\
3393&2002-05-28  05:34:44&157.4&       266.41992& -29.00407&      75.5\\
3665&2002-06-03  01:24:37& 89.3&       266.41992& -29.00407&      75.5\\
3549&2003-06-19  18:28:55& 23.8&       266.42095& -29.01052&     346.8\\
4683&2004-07-05  22:33:11& 49.5&       266.41606& -29.01240&     286.2\\
4684&2004-07-06  22:29:57& 49.2&       266.41597& -29.01239&     285.4\\
6113&2005-02-27  06:26:04&  4.9&       266.41870& -29.00350&      90.6\\
5950&2005-07-24  19:58:27& 48.5&       266.41519& -29.01225&     276.7\\
5951&2005-07-27  19:08:16& 42.3&       266.41512& -29.01222&     276.0\\
5952&2005-07-29  19:51:11& 41.2&       266.41508& -29.01222&     275.5\\
5953&2005-07-30  19:38:31& 40.2&       266.41506& -29.01221&     275.3\\
5954&2005-08-01  20:16:05& 17.8&       266.41503& -29.01218&     274.9\\
6640&2006-05-03  22:26:26&  5.1&       266.41935& -29.00380&      82.8\\
6641&2006-06-01  16:07:52&  5.1&       266.42019& -29.00437&      69.7\\
6642&2006-07-04  11:01:35&  5.1&       266.41634& -29.01240&     288.4\\
6363&2006-07-17  03:58:28& 29.4&       266.41542& -29.01231&     279.5\\
6643&2006-07-30  14:30:26&  5.0&       266.41510& -29.01221&     275.4\\
6644&2006-08-22  05:54:34&  5.0&       266.41485& -29.01205&     271.7\\
6645&2006-09-25  13:50:35&  4.5&       266.41448& -29.01197&     268.3\\
6646&2006-10-29  03:28:20&  5.1&       266.41425& -29.01181&     264.4\\
7554&2007-02-11  06:16:55&  4.8&       266.41846& -29.00332&      92.6\\
7555&2007-03-25  22:56:07&  5.1&       266.41414& -29.00002&      88.0\\
7556&2007-05-17  01:05:03&  5.0&       266.41556& -28.99973&      79.5\\
7557&2007-07-20  02:27:01&  5.0&       266.42069& -29.01498&     278.4\\
7558&2007-09-02  20:19:41&  5.0&       266.41945& -29.01543&     270.5\\
7559&2007-10-26  10:04:04&  5.0&       266.41868& -29.01564&     264.8\\
9169&2008-05-05  03:53:16& 27.6&       266.41522& -28.99981&      81.7\\
9170&2008-05-06  03:00:30& 26.8&       266.41521& -28.99981&      81.7\\
9171&2008-05-10  03:18:02& 27.0&       266.41522& -28.99980&      81.7\\
9172&2008-05-11  03:36:46& 27.4&       266.41521& -28.99981&      81.7\\
9174&2008-07-25  21:50:50& 28.2&       266.42039& -29.01521&     276.4\\
9173&2008-07-26  21:20:49& 27.8&       266.42035& -29.01521&     276.2\\
10556&2009-05-18  02:19:58&112.2&       266.41566& -28.99975&      79.0\\
11843&2010-05-13  02:12:34& 78.6&       266.41539& -28.99977&      80.7\\
13016&2011-03-29  10:30:09& 17.8&       266.41431& -28.99996&      87.6\\
13017&2011-03-31  10:30:09& 17.8&       266.41435& -28.99998&      87.4\\

\hline
\\
{  ACIS-S/HETG} &&&&&\\
13850&2012-02-06  00:38:33& 59.3&       266.41369& -29.00629&      92.2\\
14392&2012-02-09  06:18:08& 57.2&       266.41369& -29.00628&      92.2\\
14394&2012-02-10  03:17:24& 17.5&       266.41367& -29.00628&      92.2\\
14393&2012-02-11  10:14:08& 41.0&       266.41369& -29.00629&      92.2\\
13856&2012-03-15  08:46:28& 38.9&       266.41368& -29.00630&      92.2\\
13857&2012-03-17  08:58:50& 39.0&       266.41369& -29.00628&      92.2\\
13854&2012-03-20  10:13:19& 22.8&       266.41368& -29.00629&      92.2\\
14413&2012-03-21  06:45:56& 14.5&       266.41367& -29.00630&      92.2\\
13855&2012-03-22  11:25:56& 19.8&       266.41369& -29.00628&      92.2\\
14414&2012-03-23  17:49:44& 19.5&       266.41366& -29.00629&      92.2\\
13847&2012-04-30  16:17:58&151.7&       266.41426& -29.00563&      76.6\\
14427&2012-05-06  20:02:07& 78.4&       266.41426& -29.00563&      76.4\\
13848&2012-05-09  12:03:55& 96.2&       266.41427& -29.00562&      76.4\\
13849&2012-05-11  03:19:47&175.4&       266.41427& -29.00563&      76.4\\
13846&2012-05-16  10:42:22& 51.2&       266.41426& -29.00562&      76.4\\
14438&2012-05-18  04:29:45& 24.8&       266.41427& -29.00561&      76.4\\
13845&2012-05-19  10:43:37&133.5&       266.41427& -29.00563&      76.4\\
14460&2012-07-09  22:34:10& 23.4&       266.41991& -29.00884&     282.3\\
13844&2012-07-10  23:12:04& 19.8&       266.41991& -29.00884&     282.3\\
14461&2012-07-12  05:49:52& 33.3&       266.41991& -29.00885&     282.3\\
13853&2012-07-14  00:38:24& 72.7&       266.41991& -29.00885&     282.3\\
13841&2012-07-17  21:07:45& 44.5&       266.41992& -29.00885&     282.3\\
14465&2012-07-18  23:24:45& 33.9&       266.41992& -29.00886&     282.3\\
14466&2012-07-20  12:38:16& 44.5&       266.41991& -29.00884&     282.3\\
13842&2012-07-21  11:53:47&154.8&       266.41991& -29.00885&     282.3\\
13839&2012-07-24  07:04:06&135.8&       266.41991& -29.00885&     282.3\\
13840&2012-07-26  20:02:58&158.8&       266.41991& -29.00885&     282.3\\
14432&2012-07-30  12:57:08& 73.3&       266.41992& -29.00885&     282.3\\
13838&2012-08-01  17:30:32& 97.6&       266.41991& -29.00885&     282.3\\
13852&2012-08-04  02:38:43&153.9&       266.41991& -29.00885&     282.3\\
14439&2012-08-06  22:18:06&110.0&       266.41960& -29.00940&     270.7\\
14462&2012-10-06  16:33:00&131.4&       266.41953& -29.00949&     268.7\\
14463&2012-10-16  00:53:35& 30.1&       266.41954& -29.00948&     268.7\\
13851&2012-10-16  18:49:52&104.7&       266.41953& -29.00949&     268.7\\
15568&2012-10-18  08:56:30& 34.9&       266.41954& -29.00950&     268.7\\
13843&2012-10-22  16:01:55&117.2&       266.41954& -29.00949&     268.7\\
15570&2012-10-25  03:31:50& 67.5&       266.41953& -29.00949&     268.7\\
14468&2012-10-29  23:43:14&143.2&       266.41954& -29.00949&     268.7\\

\hline
\\
{  ACIS-S/Non-grating} &&&&&\\
14702&2013-05-12  10:38:50& 13.7&       266.41484& -29.00586&      80.7\\
14703&2013-06-04  08:45:16& 16.8&       266.41489& -29.00588&      80.7\\
14946&2013-07-02  06:49:44& 17.9&       266.42241& -29.01387&     290.9\\
15041&2013-07-27  01:27:17& 44.5&       266.42046& -29.01490&     276.1\\
15042&2013-08-11  22:57:58& 45.1&       266.41706& -29.01561&     253.7\\
14945&2013-08-31  10:12:46& 17.3&       266.42143& -29.01450&     283.2\\
15043&2013-09-14  00:04:52& 45.1&       266.41946& -29.01524&     269.3\\
14944&2013-09-20  07:02:56& 18.2&       266.42154& -29.01443&     284.2\\
15044&2013-10-04  17:24:48& 42.7&       266.42103& -29.01465&     280.2\\
14943&2013-10-17  15:41:05& 18.2&       266.42059& -29.01489&     277.2\\
14704&2013-10-23  08:54:30& 36.3&       266.42049& -29.00932&     274.2\\
15045&2013-10-28  14:31:14& 45.1&       266.41973& -29.01515&     271.2\\
16508&2014-02-21  11:37:48& 43.1&       266.41232& -29.00086&     100.2\\
16211&2014-03-14  10:18:27& 41.7&       266.41374& -29.00619&      90.2\\
16212&2014-04-04  02:26:27& 45.0&       266.41386& -29.00605&      87.0\\
16213&2014-04-28  02:45:05& 44.1&       266.41375& -29.00620&      90.2\\
16214&2014-05-20  00:19:11& 44.8&       266.41410& -29.00577&      80.2\\
16210&2014-06-03  02:59:23& 17.0&       266.41424& -29.00562&      76.2\\
16597&2014-07-04  20:48:12& 16.5&       266.42002& -29.00855&     287.9\\
16215&2014-07-16  22:43:52& 41.5&       266.42011& -29.00859&     289.2\\
16216&2014-08-02  03:31:41& 42.7&       266.41995& -29.00900&     281.2\\
16217&2014-08-30  04:50:12& 34.2&       266.42005& -29.00878&     285.2\\
16218&2014-10-20  08:22:28& 36.0&       266.41947& -29.00971&     265.6\\
16963&2015-02-13  00:42:04& 22.4&       266.41256& -29.00375&      92.3\\
16966&2015-05-14  08:46:51& 22.4&       266.41328& -29.00329&      84.2\\
16965&2015-08-17  10:35:47& 22.7&       266.42079& -29.01191&     272.4\\
16964&2015-10-21  06:04:57& 22.6&       266.42019& -29.01231&     265.5\\
18055&2016-02-13  08:59:23& 22.7&       266.41391& -29.00775&      92.3\\
18056&2016-02-14  14:46:01& 21.8&       266.41393& -29.00774&      92.2\\
18731&2016-07-12  18:23:59& 77.2&       266.41952& -29.00767&     281.3\\
18732&2016-07-18  12:01:38& 75.4&       266.41953& -29.00808&     272.2\\
18057&2016-10-08  19:07:12& 22.7&       266.42092& -29.01167&     267.2\\
18058&2016-10-14  10:47:43& 22.4&       266.41974& -29.00933&     266.2\\
19726&2017-04-06  03:47:13& 27.6&       266.41337& -29.00658&      86.7\\
19727&2017-04-07  04:57:18& 27.5&       266.41317& -29.00614&      86.6\\
20041&2017-04-11  03:51:22& 30.9&       266.41261& -29.00489&      86.1\\
20040&2017-04-12  05:18:22& 27.2&       266.41233& -29.00429&      86.1\\
19703&2017-07-15  22:36:07& 44.1&       266.42159& -29.01160&     270.2\\
19704&2017-07-25  22:57:27& 77.8&       266.42156& -29.01048&     276.4\\
\hline

\enddata
\tablecomments{The 122 {\it Chandra} observations (45 with ACIS-I, 38 with ACIS-S/HETG and 39 with ACIS-S non-grating) of the GC. ObsIDs carried out with the same instrument are sorted by the observation time. }
\label{tab:log}
\end{deluxetable}

\begin{deluxetable}{lccccccc}
\tabletypesize{\small}
\tablewidth{-50pt}
\tablecaption{Spectral Fit Results}
\tablehead{
\colhead{Region}&
\colhead{Instrument}&
\colhead{$N_{\rm H,1}$}&
\colhead{$\Gamma_{1}$}&
\colhead{$\chi^{2} / dof$}&
\colhead{$L_{2-10}$}&
\colhead{$N_{\rm H,2}$}&
\colhead{$\Gamma_{2}$}
\\
\colhead{(1)}&
\colhead{(2)}&
\colhead{(3)}&
\colhead{(4)}&
\colhead{(5)}&
\colhead{(6)}&
\colhead{(7)}&
\colhead{(8)}
}
\startdata
All & ACIS-I & $17.6^{+5.14}_{-4.20}$&$1.40^{+0.50}_{-0.50}$  & 12.4/16& $1.96^{+0.25}_{-0.25}$&$13.1^{+6.75}_{-4.84}$& $1.62^{+0.50}_{-0.51}$  \\[5pt]\hline
All & HETG &--&$1.18^{+0.62}_{-0.64}$ &  10.6/13 & $1.99^{+0.37}_{-0.28} $&--& $1.40^{+0.62}_{-0.64}$ \\[5pt]\hline
All & ACIS-S&--& $0.62^{+0.58}_{-0.60} $&  9.7/17 &  $2.56^{+0.33}_{-0.32}$&--&$0.83^{+0.58}_{-0.61}$ \\[10pt]
\hline
\hline
\\
Near & ACIS-I &--& $0.92^{+0.53}_{-0.53}$& 20.6/18  & $1.24^{+0.17}_{-0.18}$&--& $1.14^{+0.53}_{-0.52}$\\[5pt]\hline
Far& ACIS-I &--& $2.61^{+0.92}_{-0.95}$&  8.7/11  &  $0.74^{+0.16}_{-0.17}$&--& $2.83^{+0.92}_{-0.95}$\\[5pt]\hline
Near & HETG&--& $0.20^{+0.66}_{-0.68}$&  9.0/9  &  $1.38^{+0.24}_{-0.21}$&--& $0.43^{+0.66}_{-0.68}$ \\[5pt]\hline
Far & HETG &--&$3.09^{+0.96}_{-1.03}$ & 4.1/9 & $0.83^{+0.19}_{-0.20}$&--&$3.30^{+0.96}_{-1.03}$ \\[5pt]\hline
Near& ACIS-S&--& $-0.10^{+0.63}_{-0.66}$ & 10.3/20 & $1.89^{+0.26}_{-0.26}$&--&$0.11^{+0.63}_{-0.66}$ \\[5pt]\hline
Far& ACIS-S &--& $1.92^{+1.09}_{-1.13}$&  8.3/9  & $0.74^{+0.03}_{-0.18}$&--&$2.13^{+1.09}_{-1.13}$ \\
\enddata
\tablecomments{(1) Spectral extraction region: $'$All$'$ for the whole filament; $'$Near$'$ for the half of the filament closer to Sgr A*, defined by a 3.75$\arcsec\times1.5\arcsec$ box; $'$Far$'$ for the far-half. (3)\&(7) The fixed column density, in units of $10^{22}~\rm cm^{-2}$. (4) The best-fit  photon-index of the absorbed power-law model. (6) The unabsorbed 2-10 keV luminosity given a distance of 8 kpc, in units of $10^{32}$~erg~s$^{-1}$. (8) The best-fit  photon-index of the absorbed power-law model, with the foreground dust scattering considered. Quoted errors are at 90\% confidence level.}
\label{tab:spatial}
\end{deluxetable}

\begin{deluxetable}{lcccccp{2.0cm}r}
\tabletypesize{\small}
\tablewidth{-50pt}
\tablecaption{Spectral Fit Results of Various Epochs}
\tablehead{
\colhead{No.}&
\colhead{Epoch}&
\colhead{Instrument}&
\colhead{Exposure (ks)}&
\colhead{Net Counts}&
\colhead{$\Gamma$}&
\colhead{$\chi^{2} / dof$}&
\colhead{$L_{2-10}$}
\\
\colhead{(1)}&
\colhead{(2)}&
\colhead{(3)}&
\colhead{(4)}&
\colhead{(5)}&
\colhead{(6)}&
\colhead{(7)}&
\colhead{(8)}
}
\startdata
1 & 2002.2-2002.6 & ACIS-I & $5.39\times10^{2}$ & 230 & $1.94^{+0.64}_{-0.67} $ & 5.43/9 & $2.46^{+0.40}_{-0.40} $ \\[5pt]
2 & 2004.7-2005.8 & ACIS-I & $3.05\times10^{2}$&90 & $0.74^{+1.64}_{-1.57} $ & 5.02/5  &  $1.79^{+0.59}_{-0.60}$  \\[5pt]
3 & 2008.5-2010.5 & ACIS-I &$3.57\times10^{2}$ &84 & $0.08 ^{+1.12}_{-1.23} $& 1.32/4 & $ 2.09^{+0.58}_{-0.58}$   \\[5pt]
4 & 2012.2-2012.10   & HETG  & $2.83\times10^{3}$& 381 & $1.19^{+0.62}_{-0.64} $& 9.57/13  & $1.99^{+0.27}_{-0.28} $ \\[5pt]
5 & 2013.5-2014.10 & ACIS-S &$7.68\times10^{2}$ & 290 &$ 0.79^{+0.42}_{-0.43}$ & 11.50/17  &  $2.35^{+0.23}_{-0.24} $  \\[5pt]
6 & 2015.5-2017.4 & ACIS-S & $5.45\times10^{2}$& 217 & $0.77^{+0.63}_{-0.67} $ & 8.32/7  & $2.53^{+0.33}_{-0.32} $   \\[5pt]
  \enddata
\tablecomments{(5) Net counts in the 1--9 keV band. (6) The best-fit  photon-index of the absorbed power-law model. (8) The intrinsic 2-10 keV luminosity, given a distance of 8kpc, in units of $10^{32}$~erg~s$^{-1}$. Quoted errors are at 90\% confidence level.}
\label{tab:temporal}
\end{deluxetable}

\clearpage
%\appendix
\section*{Appendix}
For ease of reference, Table A1 lists the 2--8 keV photon flux (or $3\,\sigma$ upper limit in the case of non-detection) of G359.944 in each observation, same as presented in Figure \ref{fig:lc}.
\nopagebreak
\setcounter{table}{0}
\renewcommand\thetable{A\arabic{table}}
\begin{deluxetable}{cccr}
\centering
\tabletypesize{\small}
\tablewidth{-5pt}
\tablecaption{2--8 keV photon flux of G359.944 in individual observations}
\tablecolumns{2}
\tablehead{
\colhead{ObsID}&
\colhead{$C_{\rm tot}$}&
%\colhead{$C_{b}$}&
\colhead{$C_{\rm net}$}&
\colhead{$F_{2-8}$}\\
\colhead{(1)}&
\colhead{(2)}&
\colhead{(3)}&
\colhead{(4)}
%\colhead{(5)}
}
\startdata
{  ACIS-I} &&&\\
%\hline\\
%{  ACIS-S/HETG} &&&&\\
%\hline\\
%{  ACIS-S/Non-grating}\\
242  &  21  &        5.21  &  $      0.60^{+      1.11}_{-      0.10}$\\
15611  &  25  &        9.68  &  $      0.97^{+      1.52}_{-      0.43}$\\
15612  &  8  &        3.69  &  $      0.96^{+      1.72}_{-      0.23}$\\
2951  &  13  &        8.21  &  $      2.34^{+      3.46}_{-      1.23}$\\
2952  &  9  &        6.61  &  $      1.97^{+      2.92}_{-      1.00}$\\
2953  &  12  &        8.65  &  $      2.63^{+      3.76}_{-      1.49}$\\
2954  &  8  &        2.74  &  $      0.78^{+      1.53}_{-      0.05}$\\
2943  &  38  &       23.64  &  $      2.27^{+      2.91}_{-      1.63}$\\
3663  &  31  &       14.25  &  $      1.35^{+      1.94}_{-      0.76}$\\
3392  &  115  &       56.61  &  $      1.20^{+      1.45}_{-      0.94}$\\
3393  &  147  &       75.21  &  $      1.67^{+      1.97}_{-      1.37}$\\
3665  &  66  &       32.50  &  $      1.27^{+      1.63}_{-      0.92}$\\
3549  &  10  &        1.86  &  $<      2.97$\\
4683  &  36  &       13.50  &  $      1.21^{+      1.75}_{-      0.68}$\\
4684  &  37  &       12.11  &  $      1.14^{+      1.69}_{-      0.60}$\\
6113  &  2  &        1.52  &  $      1.11^{+      2.13}_{-      0.08}$\\
5950  &  42  &       14.24  &  $      1.32^{+      1.91}_{-      0.74}$\\
5951  &  31  &       11.38  &  $      1.15^{+      1.72}_{-      0.60}$\\
5952  &  33  &       14.81  &  $      1.40^{+      1.98}_{-      0.83}$\\
5953  &  28  &        7.90  &  $      0.86^{+      1.40}_{-      0.33}$\\
5954  &  16  &        8.34  &  $      1.89^{+      2.84}_{-      0.97}$\\
6640  &  5  &        4.04  &  $      2.82^{+      4.47}_{-      1.16}$\\
6641  &  2  &        0.00  &  $<      4.54$\\
6642  &  8  &        5.13  &  $      5.11^{+      7.97}_{-      2.34}$\\
6363  &  22  &        3.33  &  $      0.62^{+      1.20}_{-      0.06}$\\
6643  &  3  &        0.13  &  $<      6.20$\\
6644  &  3  &        0.00  &  $<      5.44$\\
6645  &  2  &        0.09  &  $<      5.64$\\
6646  &  6  &        4.56  &  $      3.27^{+      5.12}_{-      1.40}$\\
7554  &  4  &        0.00  &  $<      5.98$\\
7555  &  1  &        0.04  &  $<      4.84$\\
7556  &  6  &        4.56  &  $      4.03^{+      6.38}_{-      1.64}$\\
7557  &  5  &        3.09  &  $      2.22^{+      3.86}_{-      0.58}$\\
7558  &  3  &        1.56  &  $<      6.22$\\
7559  &  2  &        0.56  &  $<      5.17$\\
9169  &  19  &       12.30  &  $      1.84^{+      2.59}_{-      1.08}$\\
9170  &  17  &        5.51  &  $      0.75^{+      1.40}_{-      0.11}$\\
9171  &  18  &       10.82  &  $      1.71^{+      2.50}_{-      0.93}$\\
9172  &  14  &        5.86  &  $      0.84^{+      1.47}_{-      0.23}$\\
9174  &  20  &        8.99  &  $      1.14^{+      1.77}_{-      0.53}$\\
9173  &  20  &       12.82  &  $      1.65^{+      2.28}_{-      1.03}$\\
10556  &  62  &       27.06  &  $      1.10^{+      1.51}_{-      0.69}$\\
11843  &  29  &        0.00  &  $<      0.80$\\
13016  &  13  &        7.74  &  $      1.56^{+      2.38}_{-      0.74}$\\
13017  &  13  &        6.30  &  $      1.28^{+      2.13}_{-      0.44}$\\
\hline\\
{  ACIS-S/HETG} &&&\\
13850  &  19  &        9.43  &  $      1.46^{+      2.20}_{-      0.73}$\\
14392  &  21  &       10.95  &  $      1.70^{+      2.49}_{-      0.92}$\\
14394  &  4  &        0.00  &  $<      4.30$\\
14393  &  11  &        1.91  &  $<      3.00$\\
13856  &  13  &        5.34  &  $      1.23^{+      2.09}_{-      0.37}$\\
13857  &  13  &        5.34  &  $      1.24^{+      2.13}_{-      0.39}$\\
13854  &  11  &        8.13  &  $      3.24^{+      4.64}_{-      1.81}$\\
14413  &  2  &        0.09  &  $<      4.40$\\
13855  &  4  &        0.00  &  $<      3.49$\\
14414  &  7  &        4.13  &  $      1.90^{+      3.18}_{-      0.64}$\\
13847  &  66  &       36.80  &  $      2.20^{+      2.74}_{-      1.67}$\\
14427  &  24  &        6.77  &  $      0.78^{+      1.36}_{-      0.22}$\\
13848  &  36  &       16.38  &  $      1.53^{+      2.17}_{-      0.91}$\\
13849  &  47  &       18.76  &  $      0.97^{+      1.37}_{-      0.57}$\\
13846  &  16  &        8.82  &  $      1.45^{+      2.17}_{-      0.74}$\\
14438  &  5  &        1.17  &  $<      3.52$\\
13845  &  45  &       20.59  &  $      1.40^{+      1.92}_{-      0.89}$\\
14460  &  6  &        3.13  &  $      1.21^{+      2.15}_{-      0.26}$\\
13844  &  2  &        0.00  &  $<      3.16$\\
14461  &  9  &        4.69  &  $      0.85^{+      1.43}_{-      0.28}$\\
13853  &  25  &       15.91  &  $      2.00^{+      2.68}_{-      1.32}$\\
13841  &  8  &        0.00  &  $<      2.06$\\
14465  &  10  &        1.86  &  $<      2.72$\\
14466  &  13  &        2.95  &  $<      3.10$\\
13842  &  37  &        5.89  &  $      0.28^{+      0.55}_{-      0.02}$\\
13839  &  47  &       21.63  &  $      1.13^{+      1.54}_{-      0.74}$\\
13840  &  43  &       18.59  &  $      1.06^{+      1.48}_{-      0.64}$\\
14432  &  19  &        6.08  &  $      0.76^{+      1.32}_{-      0.21}$\\
13838  &  40  &       24.68  &  $      2.30^{+      2.94}_{-      1.66}$\\
13852  &  43  &       18.11  &  $      1.07^{+      1.51}_{-      0.64}$\\
14439  &  35  &       15.86  &  $      1.31^{+      1.86}_{-      0.76}$\\
14462  &  36  &        9.20  &  $      0.64^{+      1.08}_{-      0.21}$\\
14463  &  14  &       10.65  &  $      3.21^{+      4.40}_{-      2.00}$\\
13851  &  31  &       12.33  &  $      1.07^{+      1.61}_{-      0.54}$\\
15568  &  12  &        8.65  &  $      2.22^{+      3.17}_{-      1.26}$\\
13843  &  42  &       23.81  &  $      1.82^{+      2.37}_{-      1.28}$\\
15570  &  20  &        6.12  &  $      0.82^{+      1.44}_{-      0.22}$\\
14468  &  38  &       11.68  &  $      0.74^{+      1.17}_{-      0.32}$\\
\hline\\
{  ACIS-S/Non-grating}&&&\\
14702  &  13  &        5.34  &  $      1.30^{+      2.27}_{-      0.35}$\\
14703  &  19  &        7.51  &  $      1.53^{+      2.49}_{-      0.60}$\\
14946  &  18  &        7.95  &  $      1.52^{+      2.41}_{-      0.65}$\\
15041  &  38  &        8.80  &  $      0.68^{+      1.17}_{-      0.20}$\\
15042  &  44  &       17.68  &  $      1.34^{+      1.91}_{-      0.78}$\\
14945  &  14  &        3.95  &  $      0.78^{+      1.49}_{-      0.09}$\\
15043  &  47  &        7.27  &  $      0.55^{+      1.04}_{-      0.07}$\\
14944  &  13  &        5.34  &  $      1.01^{+      1.73}_{-      0.32}$\\
15044  &  43  &       12.85  &  $      1.02^{+      1.60}_{-      0.44}$\\
14943  &  18  &        8.43  &  $      1.59^{+      2.48}_{-      0.73}$\\
14704  &  41  &       18.50  &  $      1.76^{+      2.46}_{-      1.09}$\\
15045  &  47  &       24.50  &  $      1.85^{+      2.43}_{-      1.28}$\\
16508  &  35  &        9.63  &  $      0.82^{+      1.36}_{-      0.30}$\\
16211  &  50  &       27.03  &  $      2.23^{+      2.87}_{-      1.58}$\\
16212  &  29  &        0.76  &  $<      1.46$\\
16213  &  35  &       11.07  &  $      0.79^{+      1.29}_{-      0.30}$\\
16214  &  39  &       16.03  &  $      1.23^{+      1.77}_{-      0.69}$\\
16210  &  26  &       14.99  &  $      3.03^{+      4.18}_{-      1.88}$\\
16597  &  20  &       13.30  &  $      2.78^{+      3.79}_{-      1.76}$\\
16215  &  37  &       15.94  &  $      1.33^{+      1.90}_{-      0.76}$\\
16216  &  44  &       20.07  &  $      1.62^{+      2.23}_{-      1.02}$\\
16217  &  25  &       10.64  &  $      1.08^{+      1.64}_{-      0.52}$\\
16218  &  29  &       12.25  &  $      1.14^{+      1.75}_{-      0.54}$\\
16963  &  18  &        8.43  &  $      1.34^{+      2.12}_{-      0.58}$\\
16966  &  21  &        6.64  &  $      1.00^{+      1.81}_{-      0.21}$\\
16965  &  20  &        8.03  &  $      1.23^{+      1.98}_{-      0.51}$\\
16964  &  29  &       15.12  &  $      2.30^{+      3.22}_{-      1.40}$\\
18055  &  100  &        5.71  &  $<      5.62$\\
18056  &  98  &        1.80  &  $<      5.33$\\
18731  &  85  &       28.52  &  $      1.28^{+      1.76}_{-      0.81}$\\
18732  &  77  &       36.80  &  $      1.68^{+      2.14}_{-      1.23}$\\
18057  &  12  &        1.47  &  $<      2.10$\\
18058  &  20  &        6.60  &  $      0.96^{+      1.68}_{-      0.25}$\\
19726  &  17  &        3.12  &  $<      2.11$\\
19727  &  22  &        7.64  &  $      0.89^{+      1.54}_{-      0.26}$\\
20041  &  28  &       16.51  &  $      1.92^{+      2.63}_{-      1.21}$\\
20040  &  23  &       11.99  &  $      1.56^{+      2.32}_{-      0.82}$\\
19703  &  38  &       18.38  &  $      1.44^{+      1.99}_{-      0.91}$\\
19704  &  78  &       37.80  &  $      1.69^{+      2.13}_{-      1.25}$\\
\hline
\enddata
\label{tab:del_src}
\tablecomments{(1) Observation ID; (2)-(3) 2--8 keV total counts and net counts of G359.944. (4) The 2--8 keV photon flux, in units of $10^{-6}$ erg~s$^{-1}$~cm$^{-2}$. For the observations with sufficient net counts, the quoted errors are at 1-$\sigma$ confidence level. For other observations with limited net counts, 3-$\sigma$ upper limits are given (arrows in Figure \ref{fig:lc}).
}
\end{deluxetable}


\begin{thebibliography}{}
\bibitem[Ahnen et al.(2017)]{Ahnen2017} Ahnen, M.~L., Ansoldi, S., Antonelli, L.~A., et al.\ 2017, \aap, 601, A33 
\bibitem[Baganoff et al.(2001)]{Baganoff2001} Baganoff, F.~K., Bautz, M.~W., Brandt, W.~N., et al.\ 2001, \nat, 413, 45 
\bibitem[Ball et al.(2016)]{Ball2016} Ball, D., {\"O}zel, F., Psaltis, D., \& Chan, C.-k.\ 2016, \apj, 826, 77
%{   \bibitem[Blandford \& Payne(1982)]{Blandford1982} Blandford, R.~D., \& Payne, D.~G.\ 1982, \mnras, 199, 883  }
\bibitem[Broderick et al.(2016)]{Broderick2016} Broderick, A.~E., Fish, V.~L., Johnson, M.~D., et al.\ 2016, \apj, 820, 137 
%\bibitem[Burkert et al.(2012)]{Burkert2012} Burkert, A., Schartmann, M., Alig, C., et al.\ 2012, \apj, 750, 58 
\bibitem[Chan et al.(2015)]{Chan2015} Chan, C.-k., Psaltis, D., {\"O}zel, F., et al.\ 2015, \apj, 812, 103 
\bibitem[Drury(1983)]{Drury1983} Drury, L.~O.\ 1983, Reports on Progress in Physics, 46, 973 
\bibitem[Eatough et al.(2013)]{Eatough2013} Eatough, R.~P., Falcke, H., Karuppusamy, R., et al.\ 2013, \nat, 501, 391 
\bibitem[Ekers et al.(1983)]{Ekers1983} Ekers, R.~D., van Gorkom, J.~H., Schwarz, U.~J., \& Goss, W.~M.\ 1983, \aap, 122, 143
\bibitem[Falcke et al.(1993)]{Falcke1993} Falcke, H., Mannheim, K., \& Biermann, P.~L.\ 1993, \aap, 278, L1 
\bibitem[Falcke \& Markoff(2000)]{Falcke2000} Falcke, H., \& Markoff, S.\ 2000, \aap, 362, 113 
\bibitem[Gaensler \& Slane(2006)]{Gaensler2006} Gaensler, B.~M., \& Slane, P.~O.\ 2006, \araa, 44, 17 
\bibitem[Genzel et al.(2003)]{Genzel2003} Genzel, R., Sch{\"o}del, R., Ott, T., et al.\ 2003, \nat, 425, 934 
\bibitem[Ghez et al.(2008)]{Ghez2008} Ghez, A.~M., Salim, S., Weinberg, N.~N., et al.\ 2008, \apj, 689, 1044 
\bibitem[Ghez et al.(2004)]{Ghez2004} Ghez, A.~M., Wright, S.~A., Matthews, K., et al.\ 2004, \apjl, 601, L159
\bibitem[Gillessen et al.(2009)]{Gillessen2009} Gillessen, S., Eisenhauer, F., Trippe, S., et al.\ 2009, \apj, 692, 1075 
\bibitem[Gillessen et al.(2012)]{Gillessen2012} Gillessen, S., Genzel, R., Fritz, T.~K., et al.\ 2012, \nat, 481, 51
\bibitem[Gregory \& Loredo(1992)]{Gregory1992} Gregory, P.~C., \& Loredo, T.~J.\ 1992, \apj, 398, 146 
\bibitem[Johnson et al.(2009)]{Johnson2009} Johnson, S.~P., Dong, H., \& Wang, Q.~D.\ 2009, \mnras, 399, 1429 
\bibitem[Kargaltsev et al.(2017)]{Kargaltsev2017} Kargaltsev, O., Klingler, N., Chastain, S., \& Pavlov, G.~G.\ 2017, Journal of Physics Conference Series, 932, 012050  
\bibitem[Li et al.(2017)]{Li2017} Li, Y.-P., Yuan, F., \& Wang, Q.~D.\ 2017, \mnras, 468, 2552
\bibitem[Li, Morris \& Baganoff(2013)]{Li2013} Li, Z., Morris, M.~R., \& Baganoff, F.~K.\ 2013, \apj, 779, 154 (LMB13)
\bibitem[Lo \& Claussen(1983)]{Lo1983} Lo, K.~Y., \& Claussen, M.~J.\ 1983, \nat, 306, 647 
\bibitem[Lu et al.(2008)]{Lu2008} Lu, F.~J., Yuan, T.~T., \& Lou, Y.-Q.\ 2008, \apj, 673, 915 
\bibitem[Markoff et al.(2001)]{Markoff2001} Markoff, S., Falcke, H., Yuan, F., \& Biermann, P.~L.\ 2001, \aap, 379, L13 
\bibitem[Meier(2012)]{Meier2012} Meier, D.~L.\ 2012, Black Hole Astrophysics: The Engine Paradigm, by David L.~Meier.~ISBN: 978-3-642-01935-7.~Springer, Verlag Berlin Heidelberg, 2012,
\bibitem[Melia \& Falcke(2001)]{Melia2001} Melia, F., \& Falcke, H. 2001, ARA\&A, 39, 309
\bibitem[Morris et al.(2014)]{Morris2014} Morris, M.~R., Zhao, J.-H., \& Goss, W.~M.\ 2014, The Galactic Center: Feeding and Feedback in a Normal Galactic Nucleus, 303, 369 
\bibitem[Mossoux \& Grosso(2017)]{Mossoux2017} Mossoux, E., \& Grosso, N.\ 2017, \aap, 604, A85 
\bibitem[Muno et al.(2008)]{Muno2008} Muno, M.~P., Baganoff, F.~K., Brandt, W.~N., Morris, M.~R., \& Starck, J.-L.\ 2008, \apj, 673, 251 
\bibitem[Nandra et al.(1997)]{Nandra1997} Nandra, K., George, I.~M., Mushotzky, R.~F., Turner, T.~J., \& Yaqoob, T.\ 1997, \apj, 476, 70 
\bibitem[Nowak et al.(2012)]{Nowak2012} Nowak, M.~A., Neilsen, J., Markoff, S.~B., et al.\ 2012, \apj, 759, 95 
\bibitem[Pacholczyk(1970)]{Pacholczyk1970} Pacholczyk, A.~G.\ 1970, Radio Astrophysics: Nonthermal Processes in Galactic and Extragalactic Sources. Freeman \& Co., San Francisco 
\bibitem[Park et al.(2015)]{Park2015} Park, J.-H., Trippe, S., Krichbaum, T.~P., et al.\ 2015, \aap, 576, L16
%\bibitem[Pfuhl et al.(2015)]{Pfuhl2015} Pfuhl, O., Gillessen, S., Eisenhauer, F., et al.\ 2015, \apj, 798, 111 
\bibitem[Plante et al.(1995)]{Plante1995} Plante, R.~L., Lo, K.~Y., \& Crutcher, R.~M.\ 1995, \apjl, 445, L113
\bibitem[Ponti et al.(2015)]{Ponti2015} Ponti, G., De Marco, B., Morris, M.~R., et al.\ 2015, \mnras, 454, 1525 
\bibitem[Porquet et al.(2003)]{Porquet2003} Porquet, D., Predehl, P., Aschenbach, B., et al.\ 2003, \aap, 407, L17
\bibitem[Predehl \& Schmitt(1995)]{Predehl1995} Predehl, P., \& Schmitt, J.~H.~M.~M.\ 1995, \aap, 293, 889 
\bibitem[Psaltis et al.(2015)]{Psaltis2015} Psaltis, D., Narayan, R., Fish, V.~L., et al.\ 2015, \apj, 798, 15 
\bibitem[Rybicki \& Lightman(1986)]{Rybicki1986} Rybicki, G.~B., \& Lightman, A.~P.\ 1986, Radiative Processes in Astrophysics, by George B.~Rybicki, Alan P.~Lightman, pp.~400.~ISBN 0-471-82759-2.~Wiley-VCH , June 1986., 400 
\bibitem[Sch{\"o}del et al.(2010)]{Schodel2010} Sch{\"o}del, R., Najarro, F., Muzic, K., \& Eckart, A.\ 2010, \aap, 511, A18
\bibitem[Schlickeiser(1984)]{Schlickeiser1984} Schlickeiser, R.\ 1984, \aap, 136, 227
\bibitem[Shahzamanian et al.(2015)]{Sha2015} Shahzamanian, B., Eckart, A., Valencia-S., M., et al.\ 2015, \aap, 576, A20  
\bibitem[Tsuboi et al.(2015)]{Tsuboi2015} Tsuboi, M., Asaki, Y., Kameya, O., et al.\ 2015, \apjl, 798, L6 
\bibitem[Turner et al.(1999)]{Turner1999} Turner, T.~J., George, I.~M., Nandra, K., \& Turcan, D.\ 1999, \apj, 524, 667 
\bibitem[Vincent et al.(2015)]{Vincent2015} Vincent, F.~H., Yan, W., Straub, O., Zdziarski, A.~A., \& Abramowicz, M.~A.\ 2015, \aap, 574, A48 
\bibitem[Wang et al.(2006)]{Wang2006} Wang, Q.~D., Lu, F.~J., \& Gotthelf, E.~V.\ 2006, \mnras, 367, 937
\bibitem[Wang et al.(2013)]{Wang2013} Wang, Q.~D., Nowak, M.~A., Markoff, S.~B., et al.\ 2013, Science, 341, 981 
\bibitem[Wilms et al.(2000)]{Wilms2000} Wilms, J., Allen, A., \& McCray, R.\ 2000, \apj, 542, 914 
\bibitem[Witzel et al.(2014)]{Witzel2014} Witzel, G., Ghez, A.~M., Morris, M.~R., et al.\ 2014, \apjl, 796, L8 
\bibitem[Witzel et al.(2018)]{Witzel2018} Witzel, G., Martinez, G., Hora, J., et al.\ 2018, \apj, 863, 15
\bibitem[Witzel et al.(2017)]{Witzel2017} Witzel, G., Sitarski, B.~N., Ghez, A.~M., et al.\ 2017, \apj, 847, 80 
\bibitem[Yuan et al.(2002)]{Yuan2002} Yuan, F., Markoff, S., \& Falcke, H.\ 2002, \aap, 383, 854 
\bibitem[Yuan \& Narayan(2014)]{Yuan2014} Yuan, F., \& Narayan, R.\ 2014, \araa, 52, 529 
\bibitem[Yuan \& Wang(2016)]{Yuan2016} Yuan, Q., \& Wang, Q.~D. 2016, MNRAS, 456, 1438
\bibitem[Yuan et al.(2018)]{Yuan2018} Yuan, Q., Wang, Q.~D., Liu, S., \& Wu, K.\ 2018, \mnras, 473, 306
%\bibitem[Yuan et al.(2017)]{Yuan2017} Yuan, Q., Feng, L., Yin, P.-F., et al.\ 2017, arXiv:1711.10989 
%\bibitem[Yusef-Zadeh et al.(2006)]{Yusef2006} Yusef-Zadeh, F., Roberts, D., Wardle, M., Heinke, C.~O., \& Bower, G.~C.\ 2006, \apj, 650, 189 
\bibitem[Yusef-Zadeh et al.(2016)]{Yusef2016} Yusef-Zadeh, F., Wardle, M., Sch{\"o}del, R., et al.\ 2016, \apj, 819, 60 
\bibitem[Zabalza(2015)]{Zabalza2015} Zabalza, V.\ 2015, 34th International Cosmic Ray Conference (ICRC2015), 34, 922 
\bibitem[Zhang et al.(2017)]{Zhang2017} Zhang, S., Baganoff, F.~K., Ponti, G., et al.\ 2017, \apj, 843, 96 
\bibitem[Zhang et al.(2014)]{Zhang2014} Zhang, S., Hailey, C.~J., Baganoff, F.~K., et al.\ 2014, \apj, 784, 6 
\bibitem[Zhao et al.(2009)]{Zhao2009} Zhao, J.-H., Morris, M.~R., Goss, W.~M., \& An, T.\ 2009, \apj, 699, 186 
\bibitem[Zhu et al.(2018)]{Zhu2018} Zhu, Z., Li, Z., \& Morris, M.~R.\ 2018, \apjs, 235, 26

\end{thebibliography}
\end{document}